\documentclass[3p, 11pt, final]{elsarticle}
\usepackage[utf8]{inputenc}
\usepackage{pdflscape}
\usepackage{longtable,array,ragged2e}
\newcolumntype{P}[1]{>{\centering\arraybackslash\hspace{0pt}}p{#1}}
 \usepackage{fullpage}
\usepackage{threeparttable}
\usepackage{xcolor}
\usepackage{caption}
\usepackage{subcaption}

\usepackage[bordercolor=gray!20,backgroundcolor=red!10,linecolor=red!50,textsize=scriptsize,textwidth=0.85in,obeyFinal]{todonotes}
\newcommand{\blue}[1]{{\color{blue}#1}}
\usepackage{fontawesome}
\usepackage{marvosym}
\usepackage{afterpage}
\usepackage{booktabs}
\usepackage{multirow}
\usepackage{graphicx}
\usepackage{lscape}
\usepackage{soul}
\usepackage{floatrow}
\newfloatcommand{capbtabbox}{table}[][\FBwidth]
\usepackage{soul}

\usepackage{natbib}
\newcommand{\done}{\todo[color=white,bordercolor=white,textcolor=green,size=\large]{\faCheckSquareO}}

\renewcommand{\blue}[1]{{#1}}

\usepackage{tikz}
\usepackage{calc}
\def\checkmark{\tikz\fill[scale=0.5](0,.35) -- (.25,0) -- (1,.7) -- (.25,.15) -- cycle;} 
\def\scalecheck{\resizebox{\widthof{\checkmark}*\ratio{\widthof{x}}{\widthof{\normalsize x}}}{!}{\checkmark}}

\usepackage{float}

\title{Domain Adaptation for Inertial Measurement Unit-based Human Activity Recognition: A Survey}
\author{Avijoy Chakma, Abu Zaher Md Faridee, Indrajeet Ghosh, Nirmalya Roy}
\address{ \quad (achakma1, faridee1, indrajeetghosh, nroy)@umbc.edu \\ Department of Information Systems\\ 
University of Maryland Baltimore County; Baltimore; USA}

\begin{document}
\begin{frontmatter}
\begin{abstract}
\done
\blue{Machine learning-based} wearable human activity recognition (WHAR) models \blue{enable the} development of various smart and connected community applications such as sleep pattern monitoring, medication reminders, cognitive health assessment, sports analytics, etc.
However, the widespread adoption of these WHAR models is impeded by their degraded performance in the presence of data distribution heterogeneities caused by the sensor placement at different body positions, inherent biases and heterogeneities across devices, and personal and environmental diversities.
Various traditional machine learning algorithms and transfer learning techniques have been proposed in the literature to address the underpinning  challenges of handling such data heterogeneities. {\it Domain adaptation} is one such transfer learning techniques that has gained significant popularity in recent literature. In this paper, we survey the recent progress of domain adaptation techniques in the Inertial Measurement Unit (IMU)-based human activity recognition area, discuss potential future directions.
\end{abstract}
\end{frontmatter}
\section{Introduction}

% Introduce IoT and big data
%The COTS IMU devices are often equipped with a plethora of motion, muscle, ambient, object sensors, etc. that enable us to accumulate multi-sensors data simultaneously. the devices to autonomously collect massive amounts of data. 
The swift development and exceptional user-feasibility of commercial off-the-shelf~(COTS) inertial measurement unit~(IMU)~devices have created a tremendous surge in smart device applications that aim to make our everyday lives easier and more automated. Most IMU devices are equipped with multi-sensors, i.e. (accelerometer, magnetometer, and gyroscope), muscle, tactile, etc.) that enable us to accumulate massive temporal data simultaneously. As a result, the compiled multi-sensors have contributed to the development of novel data-driven machine-learning techniques in wearable human activity recognition~(WHAR). In addition, the collected multi-sensor data often contains the latent characteristics of our everyday personal events and activities, which can be effortlessly learned and predicted by sophisticated machine learning algorithms. This results in the development of various ubiquitous computing applications such as elder caring~\cite{van2010activity}, monitoring and predicting diseases~\cite{alberdi2018smart, kubota2016machine}, analyzing sports activities~\cite{wang2018inertial, chakma2020shoot, ghosh2022decoach}, developing smart home environments~\cite{wen2016adaptive}, and many more.

%Technological advancement and availability of COTS IMU devices contributed to developing novel data-driven machine learning techniques in the wearable human activity recognition~(WHAR). The IMU devices comprise a 3-axial of accelerometer, magnetometer and gyroscope sensors to estimate and capture the human performing activities' orientation, rotation and motion patterns information. 
Because of privacy concerns, IMU-based sensors have proven trustworthy compared to other modalities for human activity recognition and learning the distinctive and generalized patterns of human actions and behaviors. Human activity recognition~(HAR) recently has been a heavily investigated topic in the past decade, especially after transfer learning, deep learning, etc. techniques gained popularity due to their ability to learn the representation from the high-dimension sparse dataset. However, the explosion in the amount of collectable data and their potential application in ubiquitous applications has raised concerns about the scalability of the machine learning (more recently, deep learning) approaches used to build such applications. Furthermore, the performance of most machine learning models is heavily reliant on how much the data samples during the inference phase (i.e. test samples) match the distribution of the training samples used initially to train the machine learning model, and more often than not, the distributions differ considerably. 

%\ig{Discuss basic thought on Transfer Learning and Domain Adaptation. I would suggest to add another paragraph discussing basic idea on transfer and domain adaptation. You can merge 3rd and 4th paragraph into one paragraph.}

\begin{figure}[!htb]
    \centering
    \includegraphics[width=\linewidth, trim=6cm 4.5cm 4cm 1cm, clip]{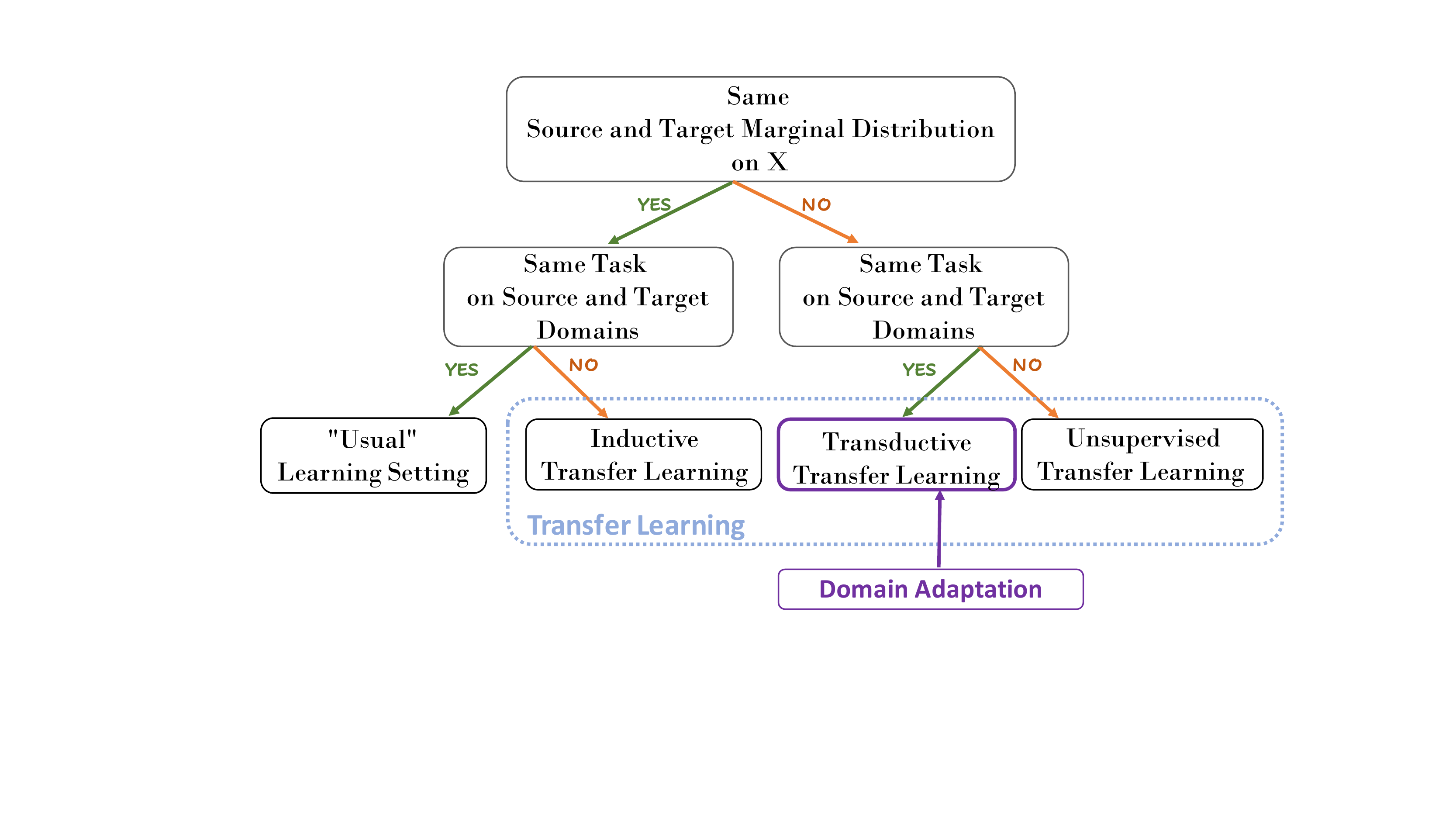}
    \caption{Marginal distribution and task similarity-based transfer learning hierarchy~\cite{redko2020survey}}
    \label{fig:redko-hierarchy}
\end{figure}

{\it Transfer learning} is a machine learning mechanism that aims to leverage a labeled dataset to learn a model for a task (i.e. classification, regression) and leverage the learned model for a similar but different task. 
There are many state-of-the-art methodologies through which transfer learning can be accomplished. Still, before we prospect down on the hierarchy of the transfer learning mechanism, we need to define two core notions: \textbf{\textit{Domain}} and \textbf{\textit{Task}}.
The term \textbf{\textit{domain}} collectively refers to the input data samples and the underlying distribution that generates them.
\textbf{\textit{Task}} is defined by a tuple that consists of a label space and the predictor function~\cite{pan2009survey}. The label space defines the number of classes the associated predictor function attempts to predict/classify. Formally,~\cite{pan2009survey} refers the \textbf{\textit{domain}} to as a two-tuple $\langle X, P(X)\rangle$ where $X$ and $P$ represent feature space and a marginal probability distribution of $X$ respectively. Feature space, $X$ consists of the data samples $x_1, x_2,...x_n \in X$. A \textbf{\textit{task}} is formally defined by two-tuple  $\langle Y, P(y|x) \rangle$, where $Y$ and $P(y|x)$ represents the label space and the predictor function. The predictor function is learned in a conditional probability distribution where $y \in Y$ and $x \in X$. In the predictor function $P(y|x)$, $y$, and $x$ present a single label and data sample respectively. \textbf{In short, we refer to the data samples (labeled or unlabeled) and the inherent data distribution as the domain.} As transfer learning leverages the learned knowledge of the existing labeled data source to accomplish a similar but different task on a new data source. The existing labeled and new data sources are colloquially referred to as the source and target datasets. Throughout this survey, we refer to the source and target datasets as the source and target domain (respectively)~\cite{kouw2019review, wilson2020survey, wang2018deepsurvey}. Depending on the domain and task similarity between two domains, different transfer learning settings emerge that are categorized as depicted in Figure~\ref{fig:redko-hierarchy}~\cite{redko2020survey}. Table~\ref{tab:tl_notation} tabulates the notation for different transfer learning settings.

\begin{table}
    \centering
    \begin{tabular}{|p{0.25\linewidth} | p{0.3\linewidth} | p{0.3\linewidth} |}\hline
        Settings & Domain Similarity & Task Similarity\\\hline
        \begin{tabular}[l]{@{}l@{}}
            \RaggedRight{``Usual'' Learning} \\
            \RaggedRight{Setting} \\
        \end{tabular} & 
        \begin{tabular}[l]{@{}l@{}}
            \RaggedRight{$(X_S == X_T)$} \textsc{and}\\
            \RaggedRight{$P(X_S) == P(X_T)$} \\
        \end{tabular} &
        \begin{tabular}[l]{@{}l@{}}
            \RaggedRight{$(Y_S == Y_T)$} \textsc{and}\\
            \RaggedRight{$P(Y_S|X_S) == P(Y_T|X_T)$} \\
        \end{tabular}\\\hline
    
        \begin{tabular}[l]{@{}l@{}}
            \RaggedRight{Inductive TL} \\
            % \RaggedRight{Transfer Learning} \\
        \end{tabular} & 
        \begin{tabular}[l]{@{}l@{}}
            \RaggedRight{$X_S == X_T$}  \textsc{and}\\
            \RaggedRight{$P(X_S) == P(X_T)$} \\
        \end{tabular} &
        \begin{tabular}[l]{@{}l@{}}
            \RaggedRight{$Y_S \; != Y_T$} \textsc{or} \\
            \RaggedRight{$P(Y_S|X_S) \; != P(Y_T|X_T)$} \\
        \end{tabular}\\\hline
        \begin{tabular}[l]{@{}l@{}}
            \RaggedRight{\textbf{Transductive TL}} \\
        \end{tabular} & 
        \begin{tabular}[l]{@{}l@{}}
            \RaggedRight{$X_S \; != X_T$} \textsc{or}\\
            \RaggedRight{$P(X_S) \; != P(X_T)$} \\
        \end{tabular} &
        \begin{tabular}[l]{@{}l@{}}
            \RaggedRight{$Y_S == Y_T$} \textsc{and} \\
            \RaggedRight{$P(Y_S|X_S) == P(Y_T|X_T)$} \\
        \end{tabular}\\\hline
    
        \begin{tabular}[l]{@{}l@{}}
            \RaggedRight{Unsupervised TL} \\
        \end{tabular} & 
        \begin{tabular}[l]{@{}l@{}}
            \RaggedRight{$X_S \; != X_T$} \textsc{or}\\
            \RaggedRight{$P(X_S) \; != P(X_T)$} \\
        \end{tabular} &
        \begin{tabular}[l]{@{}l@{}}
            \RaggedRight{$Y_S \; != Y_T$} \textsc{or} \\
            \RaggedRight{$P(Y_S|X_S) \; != P(Y_T|X_T)$} \\
        \end{tabular}\\\hline
    \end{tabular}
    \caption{Notational tabulation of different transfer learning settings. (TL refers Transfer Learning)}
    \label{tab:tl_notation}
\end{table}

Deep learning methods often result in a drop in performance under such heterogeneous data distribution scenarios. To mitigate this shortcoming, researchers have started to increasingly rely on \emph{transfer learning} techniques that leverage the learned knowledge (which is learned to accomplish a specific task, often referred to as the \textit{source task}) to accomplish a similar but different task (referred to as the \textit{target task}). Here, the machine learning model that is trained for the source task is referred to as the ``Pre-trained model''. In a typical transfer learning scenario, a pre-trained model is either partially or fully fine-tuned or re-trained, respectively~\cite{stisen2015smart, hammerla2016deep}. In the deep learning-based pre-trained model, the earlier feature extracting layers are assumed to capture the generic features embedding (consider the lines, curves, colors in object recognition task in Computer Vision) that helps in the layer layers to perform the task (example: classification/regression). Often, these approaches require a small fraction of labeled data\footnote{https://cs231n.github.io/transfer-learning/} for re-training the pre-trained model such that it performs well for the target task. However, data annotation might not always be ideal, and annotating even a fraction of data labels can be costly and cumbersome. To circumvent the data labeling problem, a particular branch of the transfer learning approach known as \textit{``Domain Adaptation''} is widely practiced in literature. According to the categorization, domain adaptation is a transfer learning setting where the marginal distributions of two domains are different even though the intended task is identical. However, we also acknowledge that different assumptions~\cite{rakshit2020multi, hu2020discriminative} on the source and target domain label space and target domain label availability have generated different domain adaptation variants, which we enumerate in Table~\ref{tab:da_type_definition}. While not comprehensive, these variations demonstrate the challenging scenarios arising from heterogeneity in the feature and label embedding spaces.

% In general, the primary causes of distribution heterogeneity are due to the variation of the user, body position, device, dataset, and data modality variability. Such change in the data distribution between the training and testing setup is known as \textit{``Domain Shift,''}~\cite{tzeng2017adversarial}. There are a few potential reasons that cause the distribution heterogeneity, such as (i) device-placement variations (e.g., hand, chest, ankle), (ii) user behavioral traits (e.g., activity execution manner, different age groups), (iii) sensor-data acquisition methodology difference among different device manufacturers.

% Table~\ref{tab:scenario_mapping} tabulates with examples, and we discuss more details on subsection~\ref{subsec:heterogeneous_env}. In the next section, we discuss the survey based on the data distribution alignment technique, network component, training mechanism, and tackled scenario that generates heterogeneity. Thus the trained machine learning models result in a performance drop when experiencing data samples that are different from the training data samples~\cite{mathur2019unsupervised}. 

% Discuss the transduction transfer learning

\begin{table}%[!H]
    \centering
    \begin{tabular}{p{0.35\linewidth} | p{0.6\linewidth}}
    \hline
    DA Type  & Assumption \\ \hline
    Closed-set DA~\cite{rakshit2020multi} & All the domains have the same set of classes~\cite{rakshit2020multi}  \\ \hline
    Open-set DA~\cite{rakshit2020multi} &  Presence of classes in the target domain that are absent in the source domain \\ \hline
    Partial DA~\cite{hu2020discriminative} &  Target label space is a subset of the source label space~\cite{hu2020discriminative} \\ \hline
    Universal DA~\cite{you2019universal} & 
    - Source label set and a target label set may contain a common label set and hold a private label set respectively~\cite{you2019universal}
    
    - UDA requires a model to either (1)
classify the target sample correctly if it is associated with a label in the common label set, or (2) mark it as “unknown” otherwise~\cite{you2019universal} \\ \hline
% bringing up an
% additional category gap. 
    Zero shot DA~\cite{wang2020adversarial} & Neither target data sample nor label is available for parameter learning~\cite{wang2020adversarial} \\ \hline
    Semi-supervised DA~\cite{li2021learning} & A small amount of labeled data from the target domain is available~\cite{li2021learning} \\ \hline
    Weakly Supervised DA~\cite{shu2019transferable} &  Source domain carries coarse labeling or corrupted data \\ \hline
    One shot DA~\cite{luo2020adversarial} & Only one unlabeled
target data sample is available~\cite{luo2020adversarial} \\ \hline
    Few shot DA~\cite{wang2019few} & Only a few labeled target domain data samples are available \\ \hline
    % Multi Source DA & \\ \midrule
    % Multi Target DA & \\ \midrule
    % Multi Step DA & \\ \midrule
    Incremental DA~\cite{liu2020learning} & Target data samples usually arrive in an online and continually evolving manner \\ \hline
    % posing challenges to classic domain adaptation paradigm: (1) Mainstream domain adaptation methods are tailored to stationary target domains, and can fail in non-stationary environments. (2) Since the target data arrive online, the model should also maintain competence on previous target data, i.e. adapt without forgetting.
    Heterogeneous DA & Source and target domain feature representation is heterogeneous or dissimilar \\ \hline
    % Target-agnostic DA &  Target domain is comprised of multiple sub-targets implicitly blended with each other, so that learners could not identify which sub-target each unlabeled sample belongs to.\\ \hline
    Federated DA~\cite{peng2019federated} & Aims to align the representations learned among the different source nodes with the data distribution of the target node where each node represents a domain (feature space)~\cite{peng2019federated}\\ \hline
    Source Free DA~\cite{li2020model} & Only unlabeled target data is available for adaptation of source prediction model and Source data is unavailable during adaptation
    \\ \hline
    % Continuously Indexed DA & \\ \hline
    \end{tabular}
    \caption{Different variants of domain adaptation. (DA refers to Domain Adaptation)}
    \label{tab:da_type_definition}
\end{table}

%\ig{Reviewers feedback -  The purpose of Section 2.3 is not clear. If the whole paper is on this topic, especially section 3, then what's the purpose of this short section? }

%\ig{My suggestion would be to discuss the domain shift problems and other challenges such as cross-body, cross-devices, orientation, etc. faced in IMU dataset in detail. I believe this would be better. Discuss more about DA and how DA can solve the challenges. Then the next section will be aligned particularly the methodology subsection 3.1.}

% While various such techniques have been explored in the computer vision and natural language processing fields~\cite{wilson2020survey, wang2018deepsurvey}, their application in IMU-based human activity recognition remains relatively modest. In addition, 

IMU-based activity recognition introduces more real-life practical scenarios that engender data distribution heterogeneity due to (i) device-placement variations (e.g., hand, chest, ankle), (ii) user behavioral traits (e.g., activity execution manner, different age groups), and (iii) sensor-data acquisition methodology differences among different device manufacturers. To comprehend the overall development of the domain adaptation approaches in IMU-based human activity recognition, we survey the existing IMU-based domain adaptation techniques, identify the existing gaps, and provide guidelines for future exploration. In addition, we also discover a few potential challenges within IMU-based activity recognition. The contributions of this survey are as follows:

\begin{itemize}

\item We systematically categorize the domain adaptation literature centered on human activity recognition according to a few consequential criteria, such as the mechanism of heterogeneity reduction and the heterogeneity being addressed. For the convenience of the readers, we have summarized the approaches, and datasets, and addressed heterogeneities in tabular form.

\item We highlight the existing limitations of domain adaptation approaches with respect to the datasets, evaluation process, scalability, real-time adaptation and lay out the guidelines for future improvements. We also discuss the research challenges that are inherent from the application perspective such as cross-modality adaptation, heterogeneous label, and task adaptation. We remark that a large majority of these challenging scenarios are yet to be thoroughly explored, and have the potential to drive innovative new research endeavors in sensor-based activity recognition if taken up by the research community. 

\end{itemize}

The survey paper is organized as follows - in section~\ref{sense-DA}, we discuss the current progress in sensor-based domain adaptation approaches based on data distribution alignment techniques and scenarios that generate heterogeneity. We provide the future guidelines and challenges of domain adaptation from an IMU-based activity recognition perspective in section~\ref{guidelines} . Finally, we conclude the survey in Section~\ref{conclusion}.

% Section~\ref{background}, we formally introduce domain adaptation, and types of data distribution heterogeneity, and follow up with a thorough literature review of the former. 

% and Section~\ref{open_research} respectively
\section{Survey}
\label{sense-DA}
In this section, we discuss the literature of IMU-based domain adaptation based on the adopted methodology and the considered heterogeneous environment in detail.

\subsection{Methodology}
Several categories of domain adaptation techniques have been proposed in the IMU-based domain adaptation literature, and these approaches mostly fall under aligning domain features, leveraging statistical normalization, inter-domain transformation of the data samples, and ensembling mechanisms. In the following sections, we discuss each of these categories and the literature associated with them.

\subsubsection{Domain Invariant Feature Learning}
\label{sub:domain_invariant}
The goal of the domain invariant feature learning technique is to align the data distribution of both source and target domains as a means of reducing the domain gap. Domain invariant feature refers to the common features from the source and target domain that are robust to the data distribution heterogeneity and the features contain task-relevant (classification/regression) information. A very common approach is to train a model using the labeled source domain and adapt the trained model using the target domain. This adaptation is guided via measuring the distribution divergence using a divergence metric, enforcing the model to learn the domain invariant features as well is maintaining the class boundaries among different classes, reducing the distribution gap through statistical normalization.

\textbf{Divergence:} Divergence is a statistical scoring mechanism of how one data distribution is similar to or different from another distribution. In domain adaptation problem scenario, divergence is used to measure how the target domain data distribution is different from the source domain data distribution. Statistically different divergence measures such as KL-Divergence~\cite{mackay2003information}, and JSD-Divergence~\cite{fuglede2004jensen} are used. Similar to the divergence, another approach to measuring the distribution difference is MMD distance~\cite{gretton2012kernel}. Here, note that MMD distance can be considered a means of divergence, but not all the divergence measures are considered as distance because the divergence measures do not always satisfy the condition of symmetry and triangle equality. KL-Divergence~\cite{mackay2003information} is not symmetric, where one data distribution acts as a reference data distribution and the distribution difference against it by the second data distribution, not the other way around. In divergence minimization-based approaches, domain adaptation is achieved by reducing the calculated divergence score between the source and target domain features~\cite{wang2018deep, wang2018stratified, khan2018scaling, faridee2019augtoact, rahman2022enabling, zhao2020local, zhao2020local, akbari2019transferring}. 
TNNAR~\cite{wang2018deep} reduces the domain gap by adapting fully connected layer via reducing the (MMD) distance between the fully-connected layer computed features of the source and target domain data whereas HDCNN~\cite{khan2018scaling} deploys separate feature extractors for the source and target domain and adapts both the Convolutional Neural Network (CNN) layers and the fully connected layer by reducing the KL-divergence. STL~\cite{wang2018stratified}, minimizes MMD distance between the feature distribution of the source domain and pseuo-labeled target domain features (more on pseudo-labeling in subsection~\ref{sub:pseudo_labeling}). Whereas AugToAct~\cite{faridee2019augtoact} deploys different feature extractors for the source and target domain and achieves the adaptation via minimizing the Jensen-Shannon Divergence~\cite{fuglede2004jensen} between extracted features of the pre-trained source network and target domain dedicated network. AugToAct~\cite{faridee2019augtoact} performs better than HDCNN~\cite{khan2018scaling} which deploys a KL-Divergence metric. Jensen-Shannon Divergence should result in stable performance for both the source and target domains because it measures the divergence of one probability distribution (P) from another (Q) in a bi-directional manner. In aligning CNN layers for domain adaptation, we believe that as the earlier CNN layers capture the domain-invariant features, the later layers should be aligned.

% \label{subsub:adversarial}
\textbf{Adversarial:} The adversarial learning mechanism attempts to align the feature distribution of the source and target domains using a domain discriminator component. The domain discriminator component can be viewed as a 2-class classifier that processes the deep extracted features from both the source and target domains and aims to predict the domain origin of the incoming features~\cite{faridee2022strangan, zhou2020xhar, chakma2021activity, hu2022swl, ragab2022self}. The goal of the domain discriminator is to correctly predict the feature origin whereas the feature extractor aims to negate the domain discriminator's capability by generating a domain invariant feature representation~\cite{ganin2015unsupervised, pei2018multi}. Contrast to the general adversarial training in using the feature extractor as the feature generator, Generative Adversarial Networks (GAN), attempts to generate targeted domain data samples from a fixed-length random vector as input~\cite{chen2019motiontransformer, sanabria2021unsupervised, sanabria2021contrasgan}. The proposed XHAR distance~\cite{zhou2020xhar} performs better than MMD and CORAL loss. ~\cite{ragab2022self} aims to boost adversarial-based domain alignment by utilizing pseudo-labeled target domain samples. MotionTransformer~\cite{chen2019motiontransformer} employs a generative adversarial training mechanism to generate target domain IMU data of the walking activity. SA-GAN~\cite{soleimani2019cross} interestingly generated fake target domain samples by adding noise to the source data samples. The key difference between MotionTransformer~\cite{chen2019motiontransformer} and SA-GAN~\cite{soleimani2019cross} is that the former does not generate samples from a random noise data distribution whereas the latter utilizes noise data to generate fake data samples.~\cite{sanabria2021unsupervised, sanabria2021contrasgan} both deploys Bi-GAN architecture to accommodate the source and target sample transformation, ~\cite{sanabria2021unsupervised} combines non-parametric distribution matching mechanism and ~\cite{sanabria2021contrasgan} deploy a contrastive learning-base mechanism for better performance.

\subsubsection{Normalization Statistics}
Normalization is a statistical scaling tool that serves the purpose of removing systematic variation and reducing noise in the data. The normalization layer is often used in the deep learning network. In deep learning networks, the batch normalization~\cite{ioffe2015batch} technique standardizes the batch input in the deep learning network training process, which subsequently helps the network to converge faster~\cite{ioffe2015batch} and increases network parameter sensitivity~\cite{ioffe2015batch}. Such a mechanism has been explored to reduce the domain gap in domain adaptation~\cite{carlucci2017autodial, li2016revisiting, mazankiewicz2020incremental}. In batch normalization, mean and variance are computed over the input batch for the respective layer during each training iteration. An exponential average of these statistics over subsequent batches is computed to be used as a global estimate for mean and variance in the testing phase~\cite{mazankiewicz2020incremental}. Instead of applying the global estimates of mean and variance,~\cite{mazankiewicz2020incremental} proposes to compute target-domain specific estimates using the fully available unlabeled target test data. Therefore, both the source and target domains are normalized using their domain-specific statistics, imposing the same target distribution on the features. The intuitive idea is that different datasets have different means and variances, and if the datasets are standardized with their corresponding means and variances, then the resulting feature space would lie under a similar data distribution~\cite{carlucci2017autodial, li2016revisiting}.

\subsubsection{Semi-supervised Techniques}
\label{sub:pseudo_labeling}
A number of proposed approaches explore semi-supervised techniques to infer the label information of the unlabeled data samples. Common approaches are active learning~\cite{akbari2020personalizing, mannini2018classifier, sztyler2017online} , co-training mechanism~\cite{wang2018stratified, chen2019cross, zhao2020local, ragab2022self, feuz2017collegial}, clustering-based similarity matching~\cite{fallahzadeh2017personalization, rokni2018autonomous}. The active learning approach determines the most uncertain and informative data samples and asks the user or Oracle for label information. Co-training mechanism trains one or more classifiers using the labeled source domain and use the classifier(s) to infer the label information based on the majority voting~\cite{wang2018stratified, chen2019cross, zhao2020local}, classifier threshold. Co-training mechanism-provided labels are known as ``pseudo labels''. Pseudo-labeled data samples help in aligning the conditional distribution~\cite{van2020survey}.

Akbari et al.~\cite{akbari2020personalizing, mannini2018classifier} explore active learning techniques in a model personalization framework to identify the ground truth labels for the most informative samples.~\cite{feuz2017collegial} proposes a teacher-learner framework, PECO, where a source data trained model predicts the pseudo-label for the target domain data using a co-training mechanism.~\cite{wang2018stratified, chen2019cross} adopts a co-training mechanism where multiple classifiers are trained with the source data and used to generate pseudo-levels for the target data based on the majority voting from the classifiers.~\cite{chen2019cross} further extend~\cite{wang2018stratified} by considering multiple source domains and selecting the most relevant domain based on the proposed stratified distance that accounts for the semantic and kinetic similarity between the source and target domain data. Whereas~\cite{rokni2018autonomous} adopts a clustering-based teacher-learner learning mechanism where the source domain acts as the teacher by providing a semi-level to the target domain for the target domain data itself. In the adaptation process, source and target data are clustered, and the clusters and labels form a complete bipartite gap where the association is achieved through the Hungarian algorithm~\cite{papadimitriou1998combinatorial}.

Table~\ref{tab:summary-all} tabulates the existing literature in details based on the proposed methodology, heterogeneity, model architecture, key learning component, and experimental datasets for the readers convenience.~\textbf{Key takeaways are:}

\begin{enumerate}
    \item ``Component'' and ``Arch'' columns from the table indicate that a significant number of proposed approaches rely on the CNN layer for feature extraction and deploy a feed-forward network for overall data processing. Very limited approaches have explored the Recurrent Neural Network (RNN) and different variations that are well known to achieve strong results in time series and sequential data.

    \item Literature has explored several loss functions (``Loss'' column) that are aimed at regularizing different aspects of the learning process. Even though these loss functions are often dependent on the deployed methodology, it is important to note that the performance behavior of the combination of these loss functions has not been studied yet. For example, it is possible to deploy divergence, adversarial, reconstruction, and contrastive loss functions together in a deep-learning based framework.

    \item ``Heterogeneity'' and ``Dataset'' columns depict that the proposed approaches tackle different heterogeneity and evaluate the methodology with different sets of datasets, which makes it really difficult to compare them. Based on observation, we note a recommendation guideline for the evaluation process discussed in subsection~\ref{sub:evaluation_guidelines}.
    
\end{enumerate}

Table~\ref{table:dataset} tabulates the different properties of the existing datasets that have be experimented to evaluate the performance of the proposed approaches. We note that HAR and WISDM datasets offers meximum experimental flexibility in terms of the user. OPPORTUNITY, DSADS, and MHEATH datasets offer flexibility in terms of body position variations. We also note that the OPPORTUNITY, PAMAP2, and DSADS datasets have been widely used in the evaluation process. Future literature should consider evaluating on the Opportunity, PAMAP2, and DSADS datasets to ensure consistency in the evaluation process. 

% Certain datasets such as OPPORTUNIY, MHEALTH, PAMAP2, DSADS, HAR, WISDM and HHAR are repeatedly used in the evaluation process.
% We also note that most of these datasets cover only the simple activities such as walking, sitting, standing and the complex activities such as office, kitchen, workshop activities are omitted. 

% Given the current dataset properties, researchers can validate the methodology robustness by experimenting the dataset that offers more experimental flexibility. 

% \AC{check for the consistency in the table and description and citation = They need to be consistent. If a paper is ignored then it should be ignored totally from the paper.}

\subsection{Heterogeneity}
\label{subsec:heterogeneous_env}
\begin{figure*}[ht]
    \centering
    \begin{subfigure}[t]{.22\linewidth}
    \includegraphics[width=\linewidth, trim=8.7cm 5.2cm 8.5cm 3cm, clip]{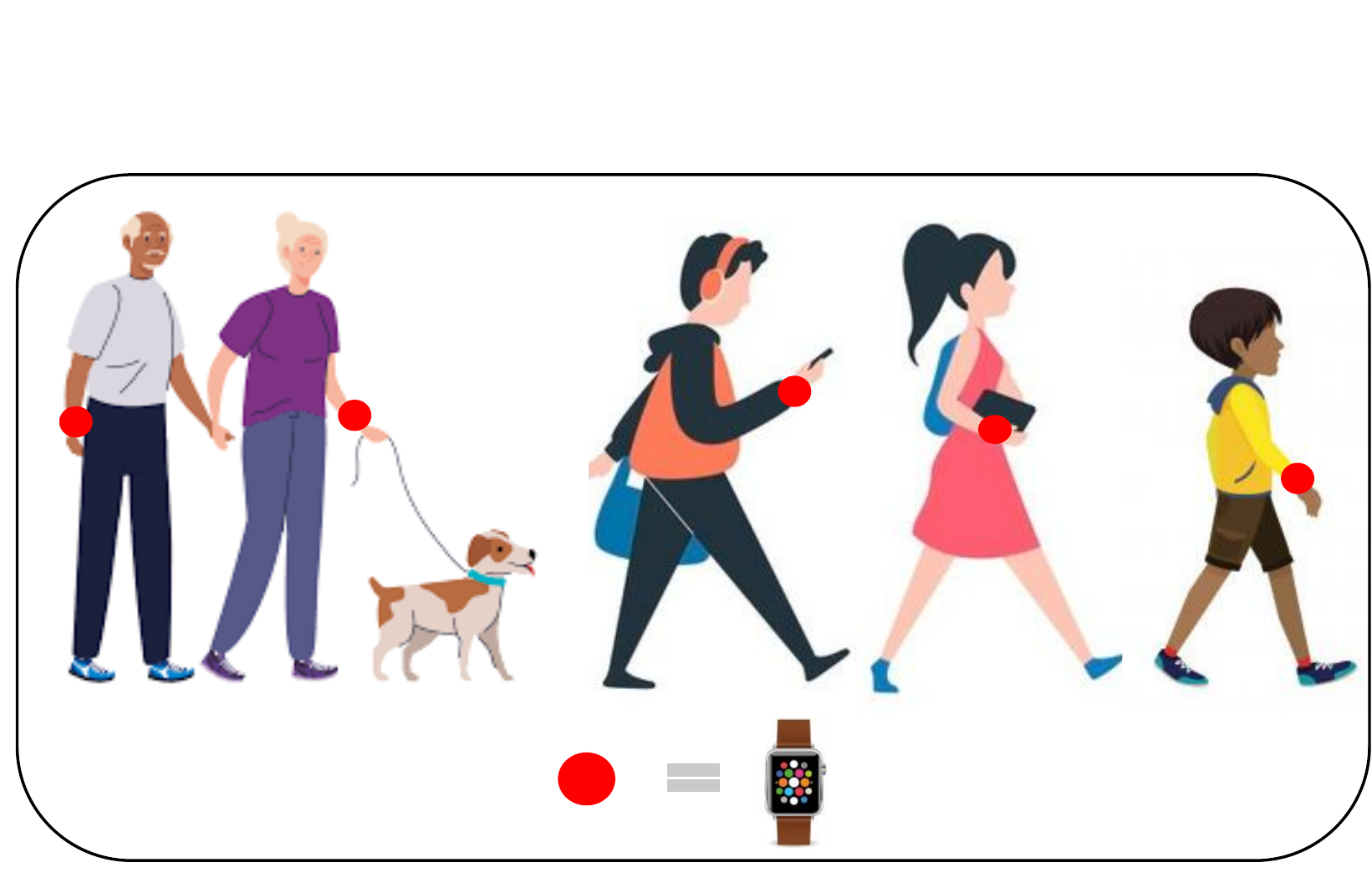}    
    \caption{Cross-person}
    \end{subfigure}
    \begin{subfigure}[t]{.25\linewidth}
    \includegraphics[width=\linewidth, trim=9.2cm 4.2cm 9cm 3cm, clip]{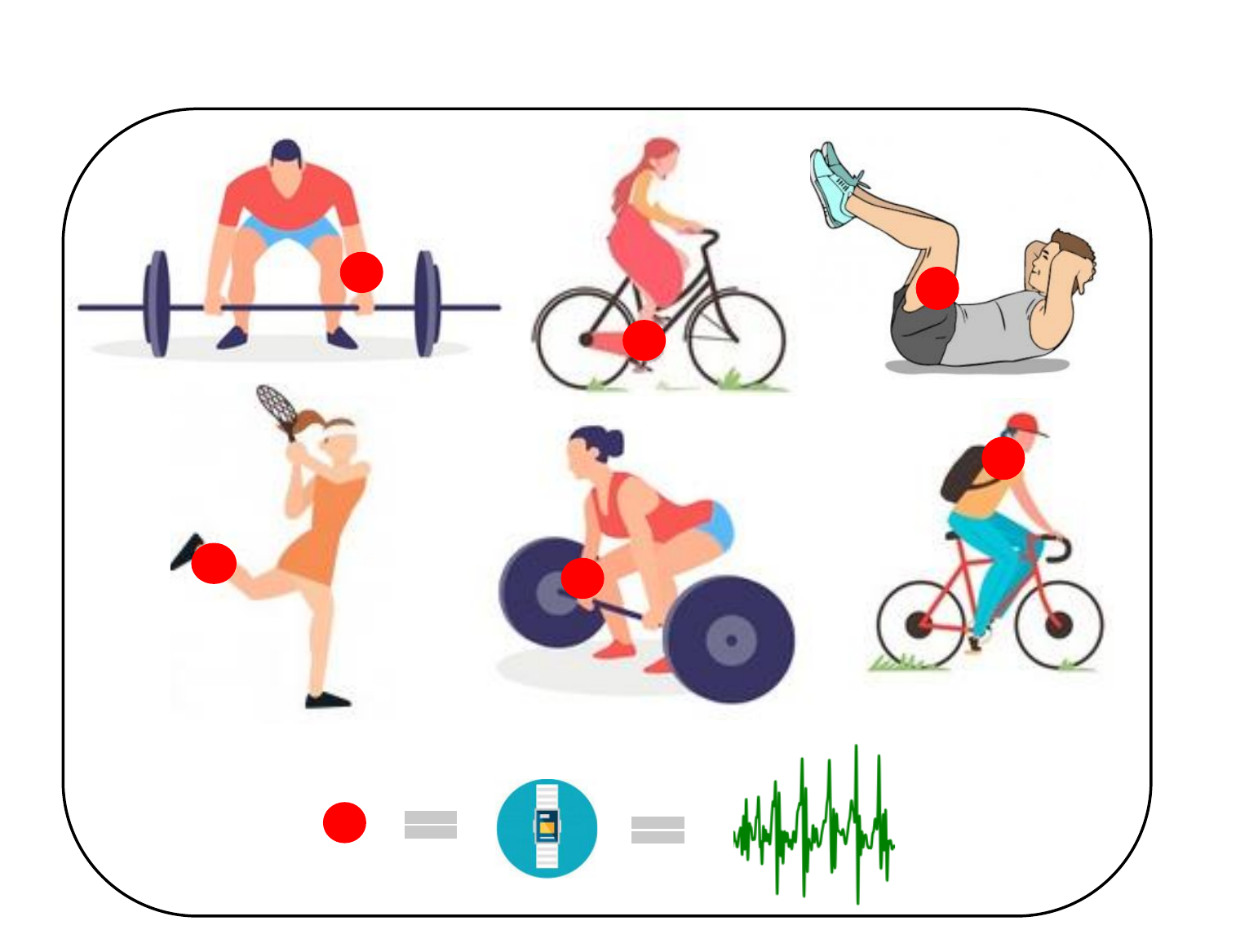}
    \caption{Cross-position}
    \end{subfigure}
    \begin{subfigure}[t]{0.5\linewidth}
    \centering
    \includegraphics[width=\linewidth, trim=6cm 7.5cm 5.5cm 5cm, clip]{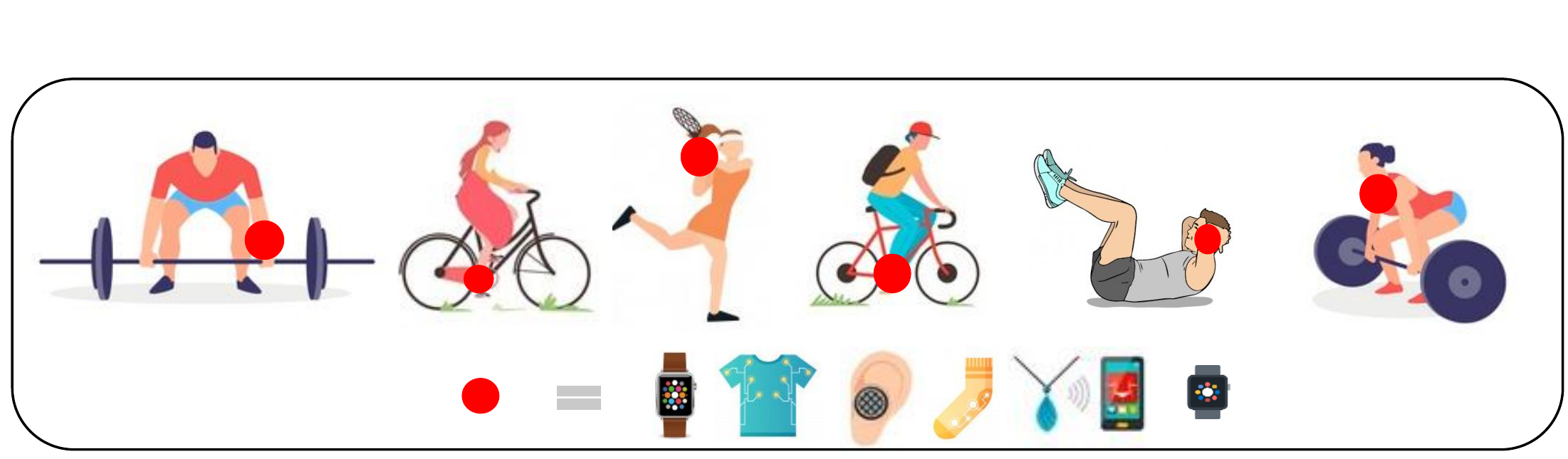}
    \caption{Cross-device}
    \end{subfigure}
    \caption{Different types of heterogeneity at a glance}
    \label{fig:various-heterogeneity}
\end{figure*}

In wearable human activity recognition, there are several practical scenarios that cause data distribution heterogeneity, such as user diversity, behavior, device placement variation, data sampling variations, and different data collection protocols for different datasets. Three commonly occurring scenarios are cross-person, cross-position, and cross-sensor/device heterogeneity, as depicted in Figure~\ref{fig:various-heterogeneity}. A example of domain adaptation settings for various heterogeneities is tabulated in Table~\ref{tab:scenario_mapping}. We discuss these heterogeneities in details in the following. 

% Some key scenarios are cross-person, cross-position, cross-device, cross-environment variation. Table~\ref{tab:heterogeneity-literature} tabulates the literature based on different heterogeneity where literature works are mentioned in one or multiple categories. Moreover, we discuss each of these with example scenario in the following.

\begin{table}[!ht]
\centering
\caption{IMU-based domain adaptation scenarios with examples.}
\label{tab:scenario_mapping}
\resizebox{.85\linewidth}{!}{%
\begin{tabular}{@{}cccc@{}}
\toprule
    Heterogeneous Scenario &
  Cause of Domain Gap &
  Source Domain &
  Target Domain \\ \midrule
Cross-person &
  Person &
  Person-A &
  Person-B \\\midrule
Cross-position &
 Position &
 Hand  &
 Ankle \\ \midrule 
Cross-dataset &
  Dataset &
  PAMAP &
  DSADS \\\midrule
Cross-device &
 Device &
 LG Nexus  &
 Samsung Galaxy \\ \midrule 
Cross-modal &
  Feature Representation &
  Acoustic &
  IMU \\\bottomrule 
% Time heterogeneity &
%  Time duration &
%  Labeled data samples  &
%  Incoming data samples \\ \bottomrule 
\end{tabular}%

}
\end{table}

% \begin{landscape}
\begin{table}[!htb]
\centering
\caption{Brief summary of IMU-based wearable domain adaptation literature.}
\label{tab:summary-all}
\resizebox{\textwidth}{!}{%
\begin{tabular}{@{}l|llllllllll|lllll|lll|llllllll|llll|ll@{}}
\toprule
\multicolumn{1}{c}{\multirow{2}{*}{Work}} &
  \multicolumn{10}{c}{Primary Method} &
  \multicolumn{5}{c}{Component} &
  \multicolumn{3}{c}{Arch} &
  \multicolumn{8}{c}{Losses/Optimizaton Criteria} &
%   \multicolumn{1}{c}{} &
  % \multicolumn{2}{c}{Alignment} &
  \multicolumn{4}{c}{Heterogenity} &
  \multicolumn{1}{c}{\multirow{2}{*}{\begin{tabular}[c]{@{}c@{}}Datasets \\ (Activities)\end{tabular}}} \\ \cmidrule(lr){2-32}
  % \multicolumn{1}{c}{} &
  %  &
  %  &
  %  &
  %  &
  %  &
  %  &
  %  &
  %  &
  %  &
  %  &
  %  &
  %  &
  %  &
  %  &
  %  &
  %  &
  %  &
  %  &
  %  &
  %  &
  %  &
  %  &
  %  &
  %  &
  %  &
  %  &
  %  &
  %  &
  %  & 
  % \multicolumn{1}{c}{}\\ \midrule 
  
\multicolumn{1}{c}{} &
  \rotatebox[origin=c]{90}{Distance Minimization} &
  \rotatebox[origin=c]{90}{Generative} &
  % \rotatebox[origin=c]{90}{Adversarial} &
  \rotatebox[origin=c]{90}{Adversarial / Generative Adversarial} &
  \rotatebox[origin=c]{90}{Contrastive} &
  \rotatebox[origin=c]{90}{Data Augmentation} &
  \rotatebox[origin=c]{90}{Adaptive Batch Normalization} &
  \rotatebox[origin=c]{90}{Co-training (Pseudo Labeling)} &
  \rotatebox[origin=c]{90}{Active Learning} &
  \rotatebox[origin=c]{90}{Clustering} &
  \rotatebox[origin=c]{90}{Subspace Alignment} &
  \rotatebox[origin=c]{90}{CNN} &
  \rotatebox[origin=c]{90}{RNN/GRU/LSTM} &
  \rotatebox[origin=c]{90}{kNN/DT/ ORF/SVM/AdaBoost/Ensemble} &
  \rotatebox[origin=c]{90}{Graphical Models} &
  \rotatebox[origin=c]{90}{Gaussian Mixture Model} &
  \rotatebox[origin=c]{90}{Feed Forward} &
  \rotatebox[origin=c]{90}{Encoder-Decoder} &
  \rotatebox[origin=c]{90}{Variational} &
  \rotatebox[origin=c]{90}{MMD/KLD/JSD} &
  \rotatebox[origin=c]{90}{Cross Entropy} &
  \rotatebox[origin=c]{90}{Adversarial} &
  \rotatebox[origin=c]{90}{Reconstruction/Regression} &
  \rotatebox[origin=c]{90}{Contrastive} &
  \rotatebox[origin=c]{90}{Cluster Association} &
  \rotatebox[origin=c]{90}{Cosine Similarity} &
  \rotatebox[origin=c]{90}{Other} &
  % \rotatebox[origin=c]{90}{Marginal} &
  % \rotatebox[origin=c]{90}{Conditional} &
  \rotatebox[origin=c]{90}{Environment} &
  \rotatebox[origin=c]{90}{Position} &
  \rotatebox[origin=c]{90}{Person} &
  \rotatebox[origin=c]{90}{Device/Sensor/Dataset} & 
  \multicolumn{1}{c}{}\\ \midrule
TNNAR~\cite{wang2018deep} &
  \scalecheck &
  \HollowBox &
  % \HollowBox &
  \HollowBox &
  \HollowBox &
  \HollowBox &
  \HollowBox &
  \HollowBox &
  \HollowBox &
  \HollowBox &
  \HollowBox &
  \scalecheck &
  \scalecheck &
  \HollowBox &
  \HollowBox &
  \HollowBox &
  \scalecheck &
  \HollowBox &
  \HollowBox &
  \scalecheck &
  \scalecheck &
  \HollowBox &
  \HollowBox &
  \HollowBox &
  \HollowBox &
  \HollowBox &
  \HollowBox &
%   \HollowBox &
  % \scalecheck &
  % \HollowBox &
  \HollowBox &
  \scalecheck &
  \HollowBox &
  \HollowBox &
  \begin{tabular}[c]{@{}l@{}}
    OPP (4),PAMAP2 (18),DSADS (19)
  \end{tabular} \\ \midrule 
MotionTransformer~\cite{chen2019motiontransformer} &
  \HollowBox &
  \HollowBox &
  % \HollowBox &
  \scalecheck &
  \HollowBox &
  \HollowBox &
  \HollowBox &
  \HollowBox &
  \HollowBox &
  \HollowBox &
  \HollowBox &
  \HollowBox &
  \scalecheck &
  \HollowBox &
  \HollowBox &
  \HollowBox &
  \HollowBox &
  \scalecheck &
  \HollowBox &
  \HollowBox &
  \HollowBox &
  \scalecheck &
  \scalecheck &
  \HollowBox &
  \HollowBox &
  \HollowBox &
  \HollowBox &
%   \HollowBox &
  % \scalecheck &
  % \HollowBox &
  \HollowBox &
  \scalecheck &
  \HollowBox &
  \HollowBox &
  \begin{tabular}[c]{@{}l@{}}
    Inertial Tracking (1)
  \end{tabular} \\ \midrule
SA-GAN~\cite{soleimani2019cross} &
  \HollowBox &
  \HollowBox &
  % \scalecheck &
  \scalecheck &
  \HollowBox &
  \HollowBox &
  \HollowBox &
  \HollowBox &
  \HollowBox &
  \HollowBox &
  \HollowBox &
  \scalecheck &
  \HollowBox &
  \HollowBox &
  \HollowBox &
  \HollowBox &
  \scalecheck &
  \HollowBox &
  \HollowBox &
  \HollowBox &
  \scalecheck &
  \scalecheck &
  \HollowBox &
  \HollowBox &
  \HollowBox &
  \HollowBox &
  \HollowBox &
%   \HollowBox &
  % \scalecheck &
  % \HollowBox &
  \HollowBox &
  \HollowBox &
  \scalecheck &
  \HollowBox &
  OPP (4) \\ \midrule
XHAR~\cite{zhou2020xhar} &
  \HollowBox &
  \HollowBox &
  % \scalecheck &
  \scalecheck &
  \HollowBox &
  \HollowBox &
  \HollowBox &
  \HollowBox &
  \HollowBox &
  \HollowBox &
  \HollowBox &
  \scalecheck &
  \scalecheck &
  \HollowBox &
  \HollowBox &
  \HollowBox &
  \scalecheck &
  \HollowBox &
  \HollowBox &
  \scalecheck &
  \scalecheck &
  \HollowBox &
  \HollowBox &
  \HollowBox &
  \HollowBox &
  \HollowBox &
  \HollowBox &
%   \HollowBox &
  % \scalecheck &
  % \HollowBox &
  \HollowBox &
  \HollowBox &
  \scalecheck &
  \scalecheck &
  \begin{tabular}[c]{@{}l@{}}
    Sport Activities (6),\\Gesture Activities (5)
  \end{tabular}
   \\ \midrule
PTN~\cite{burns2020personalized} &
  \HollowBox &
  \HollowBox &
  % \HollowBox &
  \HollowBox &
  \scalecheck &
  \HollowBox &
  \HollowBox &
  \HollowBox &
  \HollowBox &
  \HollowBox &
  \HollowBox &
  \scalecheck &
  \HollowBox &
  \scalecheck &
  \HollowBox &
  \HollowBox &
  \scalecheck &
  \HollowBox &
  \HollowBox &
  \HollowBox &
  \scalecheck &
  \HollowBox &
  \HollowBox &
  \HollowBox &
  \HollowBox &
  \HollowBox &
  \scalecheck &
%   \HollowBox &
  % \scalecheck &
  % \HollowBox &
  \HollowBox &
  \scalecheck &
  \HollowBox &
  \HollowBox &
  \begin{tabular}[c]{@{}l@{}}
    MHEALTH (12), WISDM (18), SPAR (7)
  \end{tabular}
   \\ \midrule
Hetero-DNN~\cite{mathur2018using} &
  \HollowBox &
  \scalecheck &
  % \HollowBox &
  \HollowBox &
  \HollowBox &
  \scalecheck &
  \HollowBox &
  \HollowBox &
  \HollowBox &
  \HollowBox &
  \HollowBox &
  \scalecheck &
  \HollowBox &
  \HollowBox &
  \HollowBox &
  \HollowBox &
  \scalecheck &
  \HollowBox &
  \HollowBox &
  \HollowBox &
  \scalecheck &
  \HollowBox &
  \HollowBox &
  \HollowBox &
  \HollowBox &
  \HollowBox &
  \HollowBox &
%   \HollowBox &
  % \scalecheck &
  % \HollowBox &
  \HollowBox &
  \HollowBox &
  \HollowBox &
  \scalecheck &
  \begin{tabular}[c]{@{}l@{}}
    MHEALTH (12), WISDM (18), SPAR (7)
  \end{tabular} \\ \midrule
Online DA-BN~\cite{mazankiewicz2020incremental} &
  \HollowBox &
  \HollowBox &
  % \HollowBox &
  \HollowBox &
  \HollowBox &
  \HollowBox &
  \scalecheck &
  \HollowBox &
  \HollowBox &
  \HollowBox &
  \HollowBox &
  \scalecheck &
  \HollowBox &
  \HollowBox &
  \HollowBox &
  \HollowBox &
  \scalecheck &
  \HollowBox &
  \HollowBox &
  \HollowBox &
  \scalecheck &
  \HollowBox &
  \HollowBox &
  \HollowBox &
  \HollowBox &
  \HollowBox &
  \HollowBox &
%   \HollowBox &
  % \scalecheck &
  % \HollowBox &
  \HollowBox &
  \HollowBox &
  \scalecheck &
  \HollowBox &
  WISDM (6) \\ \midrule
HDCNN~\cite{khan2018scaling} &
  \scalecheck &
  \HollowBox &
  % \HollowBox &
  \HollowBox &
  \HollowBox &
  \HollowBox &
  \HollowBox &
  \HollowBox &
  \HollowBox &
  \HollowBox &
  \HollowBox &
  \scalecheck &
  \HollowBox &
  \HollowBox &
  \HollowBox &
  \HollowBox &
  \scalecheck &
  \HollowBox &
  \HollowBox &
  \scalecheck &
  \scalecheck &
  \HollowBox &
  \HollowBox &
  \HollowBox &
  \HollowBox &
  \HollowBox &
  \HollowBox &
%   \HollowBox &
  % \scalecheck &
  % \HollowBox &
  \HollowBox &
  \HollowBox &
  \scalecheck &
  \scalecheck &
  \begin{tabular}[c]{@{}l@{}}Self Collected (8), HHAR (6),\\
  Position aware activity recognition~\cite{sztyler2016body} (8)\end{tabular} \\ \midrule
Lin et al.\cite{lin2020model} &
  \scalecheck &
  \HollowBox &
  % \HollowBox &
  \HollowBox &
  \HollowBox &
  \HollowBox &
  \HollowBox &
  \HollowBox &
  \HollowBox &
  \HollowBox &
  \HollowBox &
  \scalecheck &
  \HollowBox &
  \scalecheck &
  \HollowBox &
  \HollowBox &
  \scalecheck &
  \HollowBox &
  \HollowBox &
  \HollowBox &
  \scalecheck &
  \HollowBox &
  \HollowBox &
  \HollowBox &
  \HollowBox &
  \scalecheck &
  \HollowBox &
%   \HollowBox &
  % \scalecheck &
  % \HollowBox &
  \HollowBox &
  \HollowBox &
  \scalecheck &
  \HollowBox &
  \begin{tabular}[c]{@{}l@{}}
    DSADS (19), OPP (4), SAD (7)
  \end{tabular}
   \\ \midrule
Akbari et al.\cite{akbari2020personalizing} &
  \scalecheck &
  \scalecheck &
  % \HollowBox &
  \HollowBox &
  \HollowBox &
  \HollowBox &
  \HollowBox &
  \HollowBox &
  \scalecheck &
  \HollowBox &
  \HollowBox &
  \scalecheck &
  \HollowBox &
  \HollowBox &
  \HollowBox &
  \HollowBox &
  \HollowBox &
  \scalecheck &
  \scalecheck &
  \scalecheck &
  \scalecheck &
  \HollowBox &
  \scalecheck &
  \HollowBox &
  \HollowBox &
  \HollowBox &
  \HollowBox &
%   \HollowBox &
  % \HollowBox &
  % \HollowBox &
  \HollowBox &
  \scalecheck &
  \HollowBox &
  \scalecheck &
  \begin{tabular}[c]{@{}l@{}}
    PAMAP2 (8), MoST (8)
  \end{tabular} \\ \midrule
ActiveHARNet~\cite{gudur2019activeharnet} &
  \HollowBox &
  \HollowBox &
  % \HollowBox &
  \HollowBox &
  \HollowBox &
  \HollowBox &
  \HollowBox &
  \HollowBox &
  \scalecheck &
  \HollowBox &
  \HollowBox &
  \scalecheck &
  \HollowBox &
  \HollowBox &
  \HollowBox &
  \HollowBox &
  \scalecheck &
  \HollowBox &
  \HollowBox &
  \HollowBox &
  \scalecheck &
  \HollowBox &
  \HollowBox &
  \HollowBox &
  \HollowBox &
  \HollowBox &
  \HollowBox &
%   \HollowBox &
  % \scalecheck &
  % \scalecheck &
  \HollowBox &
  \HollowBox &
  \scalecheck &
  \HollowBox &
  \begin{tabular}[c]{@{}l@{}}
    HHAR (6), Notch (5)
  \end{tabular} \\ \midrule
UDAR~\cite{sanabria2020unsupervised} &
  \HollowBox &
  \scalecheck &
  % \HollowBox &
  \HollowBox &
  \HollowBox &
  \HollowBox &
  \HollowBox &
  \scalecheck &
  \HollowBox &
  \HollowBox &
  \HollowBox &
  \scalecheck &
  \HollowBox &
  \HollowBox &
  \HollowBox &
  \HollowBox &
  \HollowBox &
  \scalecheck &
  \scalecheck &
  \scalecheck &
  \scalecheck &
  \HollowBox &
  \HollowBox &
  \HollowBox &
  \HollowBox &
  \HollowBox &
  \HollowBox &
%   \HollowBox &
  % \HollowBox &
  % \HollowBox &
  \HollowBox &
  \HollowBox &
  \HollowBox &
  \scalecheck &
  \begin{tabular}[c]{@{}l@{}}
    CASAS (6), WSN Dataset~\cite{van2011human} (7)
  \end{tabular}
   \\ \midrule
% PLOS~\cite{jiang2018towards} &
%   \HollowBox &
%   \HollowBox &
%   % \scalecheck &
%   \scalecheck &
%   \HollowBox &
%   \HollowBox &
%   \HollowBox &
%   \HollowBox &
%   \HollowBox &
%   \HollowBox &
%   \HollowBox &
%   \scalecheck &
%   \HollowBox &
%   \HollowBox &
%   \HollowBox &
%   \HollowBox &
%   \scalecheck &
%   \HollowBox &
%   \HollowBox &
%   \HollowBox &
%   \scalecheck &
%   \scalecheck &
%   \HollowBox &
%   \HollowBox &
%   \HollowBox &
%   \HollowBox &
%   \scalecheck &
% %   \HollowBox &
%   % \HollowBox &
%   % \HollowBox &
%   \scalecheck &
%   \HollowBox &
%   \HollowBox &
%   \HollowBox &
%   \textit{Self Collected} (6) \\ \midrule
  Shift-GAN~\cite{sanabria2021unsupervised} &
  \HollowBox &
  \HollowBox &
  % \scalecheck &
  \scalecheck &
  \HollowBox &
  \HollowBox &
  \HollowBox &
  \HollowBox &
  \HollowBox &
  \HollowBox &
  \scalecheck &
  \scalecheck &
  \HollowBox &
  \scalecheck &
  \HollowBox &
  \HollowBox &
  \HollowBox &
  \scalecheck &
  \HollowBox &
  \HollowBox &
  \scalecheck &
  \scalecheck &
  \HollowBox &
  \HollowBox &
  \HollowBox &
  \HollowBox &
  \HollowBox &
%   \HollowBox &
  % \HollowBox &
  % \HollowBox &
  \HollowBox &
  \scalecheck &
  \scalecheck &
  \HollowBox &
  \begin{tabular}[c]{@{}l@{}}
    PAMAP (), DSADS (), \\ Uni. of Armsterdam ( HA, HB, HC)
  \end{tabular} \\ \midrule
  CoTMix~\cite{eldele2022cotmix} &
  \HollowBox &
  \HollowBox &
  % \scalecheck &
  \HollowBox &
  \scalecheck &
  \scalecheck &
  \HollowBox &
  \HollowBox &
  \HollowBox &
  \HollowBox &
  \HollowBox &
  \scalecheck &
  \HollowBox &
  \HollowBox &
  \HollowBox &
  \HollowBox &
  \scalecheck &
  \HollowBox &
  \HollowBox &
  \HollowBox &
  \scalecheck &
  \HollowBox &
  \HollowBox &
  \scalecheck &
  \HollowBox &
  \HollowBox &
  \HollowBox &
%   \HollowBox &
  % \HollowBox &
  % \HollowBox &
  \HollowBox &
  \HollowBox &
  \scalecheck &
  \HollowBox &
  \begin{tabular}[c]{@{}l@{}}
   SSC, UCI HAR, HHAR, WISDM
  \end{tabular} \\ \midrule
  SWL-Adapt~\cite{hu2022swl} &
  \HollowBox &
  \HollowBox &
  % \scalecheck &
  \scalecheck &
  \HollowBox &
  \HollowBox &
  \HollowBox &
  \HollowBox &
  \HollowBox &
  \HollowBox &
  \HollowBox &
  \scalecheck &
  \HollowBox &
  \HollowBox &
  \HollowBox &
  \HollowBox &
  \scalecheck &
  \HollowBox &
  \HollowBox &
  \HollowBox &
  \scalecheck &
  \scalecheck &
  \HollowBox &
  \HollowBox &
  \HollowBox &
  \HollowBox &
  \HollowBox &
%   \HollowBox &
  % \HollowBox &
  % \HollowBox &
  \HollowBox &
  \HollowBox &
  \scalecheck &
  \HollowBox &
  \begin{tabular}[c]{@{}l@{}}
    SBHAR (), OPP (), RealWorld ()
  \end{tabular} \\ \midrule
  Contras-GAN~\cite{sanabria2021contrasgan} &
  \HollowBox &
  \HollowBox &
  % \scalecheck &
  \scalecheck &
  \HollowBox &
  \HollowBox &
  \HollowBox &
  \HollowBox &
  \HollowBox &
  \HollowBox &
  \HollowBox &
  \scalecheck &
  \HollowBox &
  \HollowBox &
  \HollowBox &
  \HollowBox &
  \scalecheck &
  \HollowBox &
  \HollowBox &
  \scalecheck &
  \scalecheck &
  \scalecheck &
  \HollowBox &
  \scalecheck &
  \HollowBox &
  \HollowBox &
  \HollowBox &
%   \HollowBox &
  % \HollowBox &
  % \HollowBox &
  \HollowBox &
  \scalecheck &
  \scalecheck &
  \scalecheck &
  \begin{tabular}[c]{@{}l@{}}
    WISDM (18), DSADS (19), PAMAP (12)
  \end{tabular} \\ \midrule
  SLARDA~\cite{ragab2022self} &
  \HollowBox &
  \HollowBox &
  % \scalecheck &
  \scalecheck &
  \HollowBox &
  \HollowBox &
  \HollowBox &
  \scalecheck &
  \HollowBox &
  \HollowBox &
  \HollowBox &
  \scalecheck &
  \HollowBox &
  \HollowBox &
  \HollowBox &
  \HollowBox &
  \scalecheck &
  \HollowBox &
  \HollowBox &
  \HollowBox &
  \scalecheck &
  \scalecheck &
  \scalecheck &
  \HollowBox &
  \HollowBox &
  \HollowBox &
  \HollowBox &
%   \HollowBox &
  % \HollowBox &
  % \HollowBox &
  \HollowBox &
  \HollowBox &
  \scalecheck &
  \scalecheck &
  \begin{tabular}[c]{@{}l@{}}
   HAR, SSC, MFD
  \end{tabular} \\ \midrule
  AEDA~\cite{rahman2022enabling} &
  \scalecheck &
  \HollowBox &
  % \scalecheck &
  \scalecheck &
  \HollowBox &
  \HollowBox &
  \HollowBox &
  \HollowBox &
  \HollowBox &
  \HollowBox &
  \HollowBox &
  \scalecheck &
  \HollowBox &
  \HollowBox &
  \HollowBox &
  \HollowBox &
  \HollowBox &
  \scalecheck &
  \HollowBox &
  \scalecheck &
  \scalecheck &
  \scalecheck &
  \HollowBox &
  \HollowBox &
  \HollowBox &
  \HollowBox &
  \HollowBox &
%   \HollowBox &
  % \HollowBox &
  % \HollowBox &
  \scalecheck &
  \scalecheck &
  \HollowBox &
  \HollowBox &
  \begin{tabular}[c]{@{}l@{}}
   CASAS, PAMAP
  \end{tabular} \\ \midrule
%   IFLF~\cite{hao2021invariant} &
%   \HollowBox &
%   \HollowBox &
%   % \scalecheck &
%   \HollowBox &
%   \HollowBox &
%   \HollowBox &
%   \HollowBox &
%   \HollowBox &
%   \HollowBox &
%   \HollowBox &
%   \HollowBox &
%   \scalecheck &
%   \scalecheck &
%   \HollowBox &
%   \HollowBox &
%   \HollowBox &
%   \scalecheck &
%   \HollowBox &
%   \HollowBox &
%   \HollowBox &
%   \HollowBox &
%   \HollowBox &
%   \HollowBox &
%   \HollowBox &
%   \HollowBox &
%   \HollowBox &
% %   \HollowBox &
%   % \HollowBox &
%   % \HollowBox &
%   \HollowBox &
%   \HollowBox &
%   \scalecheck &
%   \scalecheck &
%   \begin{tabular}[c]{@{}l@{}}
%    PAMAP2, USCHAD, WISDM, MobilityAI
%   \end{tabular} \\ \midrule
  LDA~\cite{zhao2020local} &
  \scalecheck &
  \HollowBox &
  % \scalecheck &
  \HollowBox &
  \HollowBox &
  \HollowBox &
  \HollowBox &
  \scalecheck &
  \HollowBox &
  \scalecheck &
  \scalecheck &
  \HollowBox &
  \HollowBox &
  \scalecheck &
  \HollowBox &
  \HollowBox &
  \HollowBox &
  \HollowBox &
  \HollowBox &
  \scalecheck &
  \HollowBox &
  \HollowBox &
  \HollowBox &
  \HollowBox &
  \HollowBox &
  \HollowBox &
  \HollowBox &
%   \HollowBox &
  % \HollowBox &
  % \HollowBox &
  \HollowBox &
  \scalecheck &
  \scalecheck &
  \HollowBox &
  \begin{tabular}[c]{@{}l@{}}
   PAMAP, DSADS
  \end{tabular} \\ \midrule
  Xia et al.~\cite{xia2021learning} &
  \scalecheck &
  \HollowBox &
  % \scalecheck &
  \HollowBox &
  \HollowBox &
  \HollowBox &
  \HollowBox &
  \HollowBox &
  \HollowBox &
  \HollowBox &
  \HollowBox &
  \scalecheck &
  \HollowBox &
  \HollowBox &
  \HollowBox &
  \HollowBox &
  \scalecheck &
  \HollowBox &
  \HollowBox &
  \scalecheck &
  \HollowBox &
  \HollowBox &
  \scalecheck &
  \HollowBox &
  \HollowBox &
  \HollowBox &
  \HollowBox &
%   \HollowBox &
  % \HollowBox &
  % \HollowBox &
  \HollowBox &
  \HollowBox &
  \HollowBox &
  \scalecheck &
  \begin{tabular}[c]{@{}l@{}}
   PAMAP, RealWorld, OPP
  \end{tabular} \\ \midrule
  Akbari et al.~\cite{akbari2019transferring} &
  \scalecheck &
  \scalecheck &
  % \scalecheck &
  \HollowBox &
  \HollowBox &
  \HollowBox &
  \HollowBox &
  \HollowBox &
  \HollowBox &
  \HollowBox &
  \HollowBox &
  \scalecheck &
  \HollowBox &
  \HollowBox &
  \HollowBox &
  \HollowBox &
  \HollowBox &
  \scalecheck &
  \HollowBox &
  \scalecheck &
  \scalecheck &
  \HollowBox &
  \scalecheck &
  \HollowBox &
  \HollowBox &
  \HollowBox &
  \HollowBox &
%   \HollowBox &
  % \HollowBox &
  % \HollowBox &
  \HollowBox &
  \HollowBox &
  \HollowBox &
  \scalecheck &
  \begin{tabular}[c]{@{}l@{}}
    HHAR (5), MoST (10), PAMAP2 (10)
  \end{tabular} \\ \midrule
  Plug-n-learn~\cite{rokni2018autonomous} &
  \HollowBox &
  \HollowBox &
  % \scalecheck &
  \HollowBox &
  \HollowBox &
  \HollowBox &
  \HollowBox &
  \HollowBox &
  \HollowBox &
  \scalecheck &
  \HollowBox &
  \HollowBox &
  \HollowBox &
  \scalecheck &
  \HollowBox &
  \HollowBox &
  \HollowBox &
  \HollowBox &
  \HollowBox &
  \HollowBox &
  \HollowBox &
  \HollowBox &
  \HollowBox &
  \HollowBox &
  \scalecheck &
  \HollowBox &
  \HollowBox &
%   \HollowBox &
  % \HollowBox &
  % \HollowBox &
  \HollowBox &
  \HollowBox &
  \HollowBox &
  \scalecheck &
  DSADS (15) \\ \midrule
  Fallahzadeh et al.~\cite{fallahzadeh2017personalization} &
  \HollowBox &
  \HollowBox &
  % \scalecheck &
  \HollowBox &
  \HollowBox &
  \HollowBox &
  \HollowBox &
  \scalecheck &
  \HollowBox &
  \scalecheck &
  \HollowBox &
  \HollowBox &
  \HollowBox &
  \scalecheck &
  \HollowBox &
  \HollowBox &
  \HollowBox &
  \HollowBox &
  \HollowBox &
  \HollowBox &
  \scalecheck &
  \HollowBox &
  \HollowBox &
  \HollowBox &
  \scalecheck &
  \HollowBox &
  \HollowBox &
%   \HollowBox &
  % \HollowBox &
  % \HollowBox &
  \HollowBox &
  \HollowBox &
  \scalecheck &
  \HollowBox &
  DSADS (15) \\ \midrule
  STAR~\cite{abdallah2015adaptive} &
  \HollowBox &
  \HollowBox &
  % \scalecheck &
  \HollowBox &
  \HollowBox &
  \HollowBox &
  \HollowBox &
  \HollowBox &
  \scalecheck &
  \scalecheck &
  \HollowBox &
  \HollowBox &
  \HollowBox &
  \scalecheck &
  \HollowBox &
  \HollowBox &
  \HollowBox &
  \HollowBox &
  \HollowBox &
  \HollowBox &
  \scalecheck &
  \HollowBox &
  \HollowBox &
  \HollowBox &
  \HollowBox &
  \HollowBox &
  \HollowBox &
%   \HollowBox &
  % \HollowBox &
  % \HollowBox &
  \HollowBox &
  \HollowBox &
  \HollowBox &
  \scalecheck &
  \begin{tabular}[c]{@{}l@{}}
  OPP (4), WISDM (6), SPAD (4) 
  \end{tabular} \\ \midrule
  Minh et al.~\cite{minh2020improving} &
  \HollowBox &
  \HollowBox &
  % \scalecheck &
  \HollowBox &
  \HollowBox &
  \HollowBox &
  \HollowBox &
  \HollowBox &
  \scalecheck &
  \HollowBox &
  \scalecheck &
  \HollowBox &
  \HollowBox &
  \HollowBox &
  \HollowBox &
  \scalecheck &
  \HollowBox &
  \HollowBox &
  \HollowBox &
  \HollowBox &
  \scalecheck &
  \HollowBox &
  \HollowBox &
  \HollowBox &
  \HollowBox &
  \HollowBox &
  \HollowBox &
%   \HollowBox &
  % \HollowBox &
  % \HollowBox &
  \HollowBox &
  \HollowBox &
  \HollowBox &
  \scalecheck &
  PAMAP (12) \\ \midrule
  STL~\cite{wang2018stratified} &
  \scalecheck &
  \HollowBox &
  % \scalecheck &
  \HollowBox &
  \HollowBox &
  \HollowBox &
  \HollowBox &
  \scalecheck &
  \HollowBox &
  \HollowBox &
  \scalecheck &
  \HollowBox &
  \HollowBox &
  \scalecheck &
  \HollowBox &
  \HollowBox &
  \HollowBox &
  \HollowBox &
  \HollowBox &
  \scalecheck &
  \HollowBox &
  \HollowBox &
  \HollowBox &
  \HollowBox &
  \HollowBox &
  \HollowBox &
  \HollowBox &
%   \HollowBox &
  % \HollowBox &
  % \HollowBox &
  \HollowBox &
  \scalecheck &
  \HollowBox &
  \scalecheck &
  \begin{tabular}[c]{@{}l@{}}
  OPP (4), PAMAP2(18), DSADS (19) 
  \end{tabular} \\ \midrule
  Wen et al.~\cite{wen2016sensor} &
  \HollowBox &
  \HollowBox &
  % \scalecheck &
  \HollowBox &
  \HollowBox &
  \HollowBox &
  \HollowBox &
  \scalecheck &
  \HollowBox &
  \HollowBox &
  \HollowBox &
  \HollowBox &
  \HollowBox &
  \scalecheck &
  \scalecheck &
  \HollowBox &
  \HollowBox &
  \HollowBox &
  \HollowBox &
  \HollowBox &
  \HollowBox &
  \HollowBox &
  \HollowBox &
  \HollowBox &
  \HollowBox &
  \HollowBox &
  \scalecheck &
%   \HollowBox &
  % \HollowBox &
  % \HollowBox &
  \HollowBox &
  \HollowBox &
  \HollowBox &
  \scalecheck &
  \begin{tabular}[c]{@{}l@{}}
  SAD (7), HAR (6),OPP (Fine-grained) 
  \end{tabular} \\ \midrule
  Mannini et al.~\cite{mannini2018classifier} &
  \HollowBox &
  \HollowBox &
  % \scalecheck &
  \HollowBox &
  \HollowBox &
  \HollowBox &
  \HollowBox &
  \scalecheck &
  \scalecheck &
  \HollowBox &
  \HollowBox &
  \HollowBox &
  \HollowBox &
  \scalecheck &
  \HollowBox &
  \HollowBox &
  \HollowBox &
  \HollowBox &
  \HollowBox &
  \HollowBox &
  \HollowBox &
  \HollowBox &
  \HollowBox &
  \HollowBox &
  \HollowBox &
  \HollowBox &
  \scalecheck &
%   \HollowBox &
  % \HollowBox &
  % \HollowBox &
  \HollowBox &
  \HollowBox &
  \scalecheck &
  \HollowBox &
  \begin{tabular}[c]{@{}l@{}}
  Adult \& Youth \\Dataset~\cite{mannini2017activity} 
  \end{tabular} \\ \midrule
  Sztyler et al.~\cite{sztyler2017online} &
  \HollowBox &
  \HollowBox &
  % \scalecheck &
  \HollowBox &
  \HollowBox &
  \HollowBox &
  \HollowBox &
  \HollowBox &
  \scalecheck &
  \HollowBox &
  \HollowBox &
  \HollowBox &
  \HollowBox &
  \scalecheck &
  \HollowBox &
  \HollowBox &
  \HollowBox &
  \HollowBox &
  \HollowBox &
  \HollowBox &
  \HollowBox &
  \HollowBox &
  \HollowBox &
  \HollowBox &
  \HollowBox &
  \HollowBox &
  \scalecheck &
%   \HollowBox &
  % \HollowBox &
  % \HollowBox &
  \HollowBox &
  \HollowBox &
  \scalecheck &
  \HollowBox &
  ~\cite{sztyler2016body} \\ \midrule
  PECO~\cite{feuz2017collegial} &
  \HollowBox &
  \HollowBox &
  % \scalecheck &
  \HollowBox &
  \HollowBox &
  \HollowBox &
  \HollowBox &
  \scalecheck &
  \HollowBox &
  \HollowBox &
  \scalecheck &
  \HollowBox &
  \HollowBox &
  \scalecheck &
  \HollowBox &
  \HollowBox &
  \HollowBox &
  \HollowBox &
  \HollowBox &
  \HollowBox &
  \HollowBox &
  \HollowBox &
  \HollowBox &
  \HollowBox &
  \HollowBox &
  \HollowBox &
  \scalecheck &
%   \HollowBox &
  % \HollowBox &
  % \HollowBox &
  \HollowBox &
  \HollowBox &
  \HollowBox &
  \scalecheck &
  \begin{tabular}[c]{@{}l@{}}
  CASAS PUCK, CASAS Parkinson 
  \end{tabular} \\ \bottomrule
%   Plug-n-learn~\cite{rokni2018autonomous} &
%   \HollowBox &
%   \HollowBox &
%   % \scalecheck &
%   \HollowBox &
%   \HollowBox &
%   \HollowBox &
%   \HollowBox &
%   \HollowBox &
%   \HollowBox &
%   \HollowBox &
%   \HollowBox &
%   \HollowBox &
%   \HollowBox &
%   \HollowBox &
%   \HollowBox &
%   \HollowBox &
%   \HollowBox &
%   \HollowBox &
%   \HollowBox &
%   \HollowBox &
%   \HollowBox &
%   \HollowBox &
%   \HollowBox &
%   \HollowBox &
%   \HollowBox &
%   \HollowBox &
% %   \HollowBox &
%   % \HollowBox &
%   % \HollowBox &
%   \HollowBox &
%   \HollowBox &
%   \HollowBox &
%   \HollowBox &
%   \textit{Self Collected} (6) \\ \bottomrule
\end{tabular}%
}
\end{table}
% \end{landscape}

\subsubsection{Cross-person Heterogeneity} 
Cross-person heterogeneity in wearables refers to the variability in sensor data and human movement patterns across different individuals, i.e. data collected from wearable devices can vary significantly from person to person. Differences in body size, shape, movement patterns, and behavior traits cause this significant discrepancy. In the literature, the problem of cross-person heterogeneity in wearables is often referred to as model personalization. Several different approaches have been proposed to address cross-person heterogeneity. Along with deep learning-based approaches, there are several non-deep learning-based approaches for addressing cross-person heterogeneity~\cite{fallahzadeh2017personalization, mannini2018classifier, sztyler2017online}. The major drawback of~\cite{fallahzadeh2017personalization, mannini2018classifier} lies in the evaluation process. The corresponding proposed methodology is evaluated on a single dataset using the leave one subject out (LOSO) validation strategy. LOSO is more similar to the traditional machine evaluation approach and does not truly represent effectiveness in tackling the data distribution heterogeneity. 

In contrast, several deep learning-based approaches are proposed that aim to reduce the distribution heterogeneity that can be discussed under unsupervised and semi-supervised learning mechanism~\cite{faridee2019augtoact, akbari2020personalizing, wang2018deep, sanabria2021contrasgan, eldele2022cotmix}. Unsupervised approaches assume the availability of the labeled source domain sample and the unlabeled target domain sample during the adaptation process. Among the unsupervised approaches, there are approaches that attempt to measure the domain gap via a distance metric calculation between the source domain feature and target domain feature, and by reducing that gap, aim to reduce the heterogeneity~\cite{wang2018deep}. Adversarial techniques are another approach that has recently been investigated for tackling cross-person heterogeneity~\cite{soleimani2019cross, zhou2020xhar, sanabria2021contrasgan, sanabria2021unsupervised}.~\cite{soleimani2019cross} leverages a general GAN-based approach whereas~\cite{sanabria2021contrasgan, sanabria2021unsupervised} deploy a bi-directional GAN architecture in the corresponding methodology. Although these approaches share the goal of reducing cross-person heterogeneity, comparing them is difficult due to differences in the evaluation procedure. For instance, the evaluation procedure differs in terms of the considered dataset(s), number of activities, compared baselines, and experimental design. For the sake of comparison, if we consider~\cite{wang2018deep, sanabria2021contrasgan} as a representative of distance-based and adversarial-based approach where both consider DSADS and PAMAP dataset in the evaluation process, from the reported results, we observe that GAN-based approach~\cite{sanabria2021contrasgan} performs superior than the distance-matric-based approach~\cite{wang2018deep} by a significant margin. In addition, the ablation study in~\cite{sanabria2021contrasgan} shows that the ensemble of losses with the GAN loss increases the performance. Different from these works,~\cite{mazankiewicz2020incremental} tackles real-time domain adaptation using statistical batch normalization-based techniques.

On the other hand, a number of semi-supervised approaches have been proposed in the literature~\cite{faridee2019augtoact, akbari2020personalizing, khan2018scaling, eldele2022cotmix, ragab2022self, hao2021invariant, zhao2020local} that vary in terms of the target domain labeled sample usage. One of the most common approaches is to use a small amount of labeled data from the target domain together with a larger amount of labeled data from the source domain to train a model that can effectively generalize to the target domain~\cite{khan2018scaling, faridee2019augtoact, hao2021invariant, lin2020model}. This approach reduces the need for large amounts of labeled data in the target domain, which can be difficult and expensive to obtain. In AugToAct~\cite{faridee2019augtoact} proposed an data augmentation-based semi-supervised approach and AugToAct~\cite{faridee2019augtoact} performs better than~\cite{khan2018scaling} in the similar experimental settings. In both approaches~\cite{khan2018scaling, faridee2019augtoact}, 20\% labeled target domain data retains high classification performance. Self-training is another variation of semi-supervised approach that leverages labeled source domain samples to train a model, and then uses the model to predict the labels of the unlabeled target domain samples, which are often known as \textit{Pseudo-labels}~\cite{van2020survey}. Pseudo-labeled data samples are then used to further improve the model~\cite{wang2018stratified, ragab2022self, zhao2020local}. Even though~\cite{wang2018stratified, zhao2020local} both operate on the manually extracted features, from the reported results on the cross-person heterogeneity, it is apparent that the clustering-based pseudo labeling technique performs better by significant margin. In addition, active learning is also explored to reduce the data distribution heterogeneity, where a trained model (trained with labeled source data samples) is able to actively choose the examples it wants to be labeled next by the oracle (human or other source of ground truth) rather than passively relying on a fixed set of labeled examples. The model uses the information it has learned so far to identify examples that it is uncertain about and would most benefit from having labeled~\cite{akbari2020personalizing}. In comparison between active learning~\cite{akbari2020personalizing} and self-learning~\cite{zhao2020local} approach,~\cite{zhao2020local} yields more reliable performance on the commonly experimented PAMAP dataset. Even though~\cite{akbari2020personalizing} outperforms~\cite{zhao2020local} by 14.88\% but~\cite{akbari2020personalizing} trains the source model using the combined data from multiple users whereas~\cite{zhao2020local} trains the source model using a single user data. Here,~\cite{akbari2020personalizing} allows the source model to observe more data variations compared to~\cite{zhao2020local}. Compared to unsupervised approaches, semi-supervised approaches align both the marginal probability and conditional probability distributions of source and target domain data samples.

\begin{table}[hbt!]
\centering
\caption{IMU-based wearable datasets (\textbf{A}, \textbf{G}, \textbf{M} in the sensor column refers to accelerometer, gyroscope and magnetometer respectively).}
\label{table:dataset}
\resizebox{\textwidth}{!}{%
\begin{tabular}{@{}lcccrlrrl@{}}
\toprule
\multirow{2}{*}{Dataset} &
  \multicolumn{3}{c}{Sensors} &
  \multicolumn{1}{c}{\multirow{2}{*}{\begin{tabular}[c]{@{}c@{}}Sampling\\ Rate\end{tabular}}} &
  \multirow{2}{*}{\begin{tabular}[c]{@{}l@{}}Body  Position\end{tabular}} &
  \multicolumn{1}{c}{\multirow{2}{*}{User}} &
  \multicolumn{1}{c}{\multirow{2}{*}{Activities}} &
  \multirow{2}{*}{Literature} \\ \cmidrule(lr){2-4}
 &
  \rotatebox[origin=c]{00}{A} &
  \rotatebox[origin=c]{00}{G} &
  \rotatebox[origin=c]{00}{M} &
  \multicolumn{1}{c}{} &
   &
  \multicolumn{1}{c}{} &
  \multicolumn{1}{c}{} &
   \\ \midrule
OPPOTUNITY \cite{chavarriaga2013opportunity, roggen2013adarc, roggen2010collecting} &
  \scalecheck &
  \scalecheck &
  \scalecheck &
  30 &
  [BACK, RUA, RLA, LUA, LLA]\footnotemark[1] &
  4 &
  4 &
  \cite{chavarriaga2013opportunity, abdallah2015adaptive, feuz2017collegial, chen2019cross, soleimani2019cross, morales2016deep, wang2018deep, li2018domain, wang2018stratified} \\ \midrule
PAMAP2 \cite{reiss2012introducing} &
  \scalecheck &
  \scalecheck &
  \scalecheck &
  100 &
  Hand, Chest, Ankle &
  9 &
  18 &
  \cite{chen2019cross, wang2018deep, li2018domain, akbari2020personalizing, wang2018stratified, barbosa2018unsupervised} \\ \midrule
DSADS \cite{barshan2014recognizing} &
  \scalecheck &
  \scalecheck &
  \scalecheck &
  25 &
  [TORSO, RA, LA, RL, LL]\footnotemark[2] &
  8 &
  19 &
  \cite{chen2019cross, wang2018deep, li2018domain, fallahzadeh2017personalization, rokni2018personalized, wang2018stratified, khan2017transact} \\ \midrule
MHEALTH \cite{banos2014mhealthdroid} &
  \scalecheck &
  \scalecheck &
  \scalecheck &
  50 &
  Chest, Right Wrist, Left Ankle &
  10 &
  12 &
  \cite{chen2019cross, khan2017transact} \\ \midrule
HHAR \cite{stisen2015smart} &
  \scalecheck &
  \scalecheck &
  \HollowBox &
  25-200 &
  Waist, Arms &
  6 &
  9 &
  \cite{gudur2019activeharnet, khan2018scaling} \\ \midrule
HAR \cite{anguita2012human} &
  \scalecheck &
  \scalecheck &
  \HollowBox &
  50 &
  Waist &
  30 &
  6 &
  \cite{almaslukh2018robust, khan2018scaling, khan2017transact} \\ \midrule
WISDM \cite{kwapisz2011activity} &
  \scalecheck &
  \scalecheck &
  \HollowBox &
  20 &
  Pant Pocket, Waist &
  51 &
  18 &
  \cite{abdallah2015adaptive, rokni2018personalized} \\ \midrule
Susses-Huawei Dataset \cite{gjoreski2018university} &
  \scalecheck &
  \scalecheck &
  \scalecheck &
  100 &
  Torso, Backpack, Hand, Pocket &
  3 &
  8 &
  \cite{gurov2019human} \\ \midrule
SAD~\cite{shoaib2014fusion} &
  \scalecheck &
  \scalecheck &
  \scalecheck &
  50 & 
  [RP, LP, BELT, RUA, RW]\footnotemark[3] &
  10 &
  7 &
  \cite{gurov2019human} \\ \midrule
SPAD \cite{do2012healthylife} &
  \scalecheck &
  \HollowBox &
  \HollowBox &
  5 & [PP, JP, HB, SB]\footnotemark[4] &
  8 &
  4 &
  \cite{abdallah2015adaptive} \\ \midrule
Notch Dataset \cite{mauldin2018smartfall} &
  \scalecheck &
  \HollowBox &
  \HollowBox &
  31.25 &
  Wrist &
  7 &
  5 &
  \cite{gudur2019activeharnet} \\ \midrule
CASAS~\cite{cook2010learning} &
  \scalecheck &
  \HollowBox &
  \HollowBox &
  - &
  Wrist, Heap &
  10 &
  6 &
  \cite{feuz2017collegial} \\ \midrule
Everyday Activities \cite{consolvo2008activity} &
  \scalecheck &
  \HollowBox &
  \HollowBox &
  - &
  Smartphone &
  41 &
  7 &
  \cite{lane2011enabling} \\ \midrule
Smartphone Dataset\cite{shoaib2013towards} &
  \scalecheck &
  \scalecheck &
  \scalecheck &
  50 &
  Arm, Belt, Waist, Pocket &
  4 &
  6 &
  \cite{wen2016sensor} \\ \bottomrule
\end{tabular}%
}
\end{table}
\footnotetext[1]{Back, Right Upper Arm, Right Lower Arm, Left Upper Arm, Left Lower Arm}
\footnotetext[2]{Torso, Right Arm, Left Arm, Right Leg, Left Leg}
\footnotetext[3]{Right Pocket, Left Pocket, Belt, Right Upper Arm, Right Wrist}
\footnotetext[4]{Pant Pocket, Jacket Pocket, Hand Bag, Shoulder Bag}

\subsubsection{Cross-position Heterogeneity}
Cross-position heterogeneity in wearables refers to the variability in sensor readings when the sensor is placed in different positions on the body. This variability can arise from factors such as the sensor placement location, the sensor orientation, and the body movement. For example, the readings from an accelerometer placed on the wrist will be different from the readings of the same accelerometer placed on the ankle due to the different movement patterns of these body parts. To address this issue, researchers have proposed a variety of techniques to reduce the cross-position heterogeneity by aligning the sensor readings across different positions~\cite{sanabria2021unsupervised, sanabria2021contrasgan, wang2018deep, chen2019motiontransformer, zhao2020local, wang2018stratified, chen2019cross, rahman2022enabling}. Among the unsupervised approaches, a handful of Generative Adversarial Networks (GANs) have been proposed as a technique to tackle cross-position heterogeneity in wearables~\cite{sanabria2021unsupervised, sanabria2021contrasgan,chen2019motiontransformer}. In domain adaptation settings, where the generator component of the GAN architecture is trained to align the feature distributions between the source and target domains, it generates synthetic samples from the source domain that are similar to the real samples from the target domain. By training the generator network to generate synthetic readings that are similar to real readings from different sensor positions, the model can learn to reduce the cross-position heterogeneity and improve the performance of the activity recognition system when applied to different sensor positions. Among the GAN-based approaches,~\cite{chen2019motiontransformer} considers single dataset with only one activity where the device is placed at different body positions whereas~\cite{sanabria2021unsupervised, sanabria2021contrasgan} considers multiple datasets in the evaluation process. Both~\cite{sanabria2021unsupervised, sanabria2021contrasgan} leverages Bi-GAN architectures but differs in the distribution alignment component where~\cite{sanabria2021unsupervised} deploys Kernel Mean Matching (KMM)~\cite{jitkrittum2019kernel} and~\cite{sanabria2021contrasgan} leverages a combination of contrastive, discrepancy and MMD losses. Overall,~\cite{sanabria2021contrasgan} evaluates more cross-positional heterogeneity from the DSADS and PAMAP datasets and performs better than~\cite{sanabria2021unsupervised, wang2018deep}.

Compared to the unsupervised approaches, even though semi-supervised approaches leverage labeled target domain data, the reported experimental results on the similar datasets are less convincing.~\cite{zhao2020local, wang2018stratified, chen2019cross} performance on the DSADS and PAMAP dataset ranges between 40-60\%. We speculate that as these approaches generate pseudo-label for the target domain data, incorrect label predictions on the target domain sample might have contributed to negative learning and hence such relatively lower performance compared to the unsupervised approaches. In contrast,~\cite{rahman2022enabling} proposes to leverage 10\% labeled data in a encoder-decoder-based architecture that adopts Gradient Reversal Layer for data distribution alignment, achieves 92.1\% accuracy on PAMAP dataset.

\subsubsection{Cross-device Heterogeneity} 
Cross-sensor heterogeneity in wearables refers to the variability in sensor readings when different device-integrated sensors are used to measure the same physical activity. For example, if two smartphones use the same accelerometer, the sensor readings may be different due to factors such as measurement ranges, sensor manufacturer, sensor model, and sensor calibration. Note that from the experiment design perspective, cross-sensor heterogeneity is also be referred as cross-device and cross-dataset heterogeneity. Even though, in cross-dataset heterogeneity the data distribution heterogeneity between two dataset arises due to the variability such as differences in the population of subjects, differences in the environment in which the data was collected, and differences in the protocol used to collect the data. Due to the integrated nature of the WHAR, we discuss these heterogeneities together. We list the common scenarios below that could raise four types of heterogeneity:

\begin{enumerate}
    \item similar sensor but different devices~\cite{wang2018stratified, hao2021invariant, zhou2020xhar, sanabria2021contrasgan}. Example - Wrist-worn Smartwatches from multiple manufacturers carry IMU sensors. Example - (devices can be treated as source and target device), here, \textbf{body position is fixed}
    \item same device but different sensors~\cite{wen2016sensor}. Example - A general smartphone has multiple sensors (IMU and Gyroscope) that are capable of collecting data (sensors can be treated as source and target sensors), here, \textbf{body position is fixed}
    \item similar sensors but different devices~\cite{khan2018scaling, akbari2019transferring}. For example - Smartphones and smartwatches are typically carried IMU sensors but worn at different body positions. In this case, \textbf{different body positions are used.}
    \item different sensors with different devices~\cite{feuz2015transfer, feuz2017collegial, sanabria2020unsupervised}. Example - Smartphones and smart-earable are typically carried IMU and microphone sensors respectively and worn at different body positions.  In this case, \textbf{different body positions are used.}
\end{enumerate}

Note that the following heterogeneity scenarios can include/exclude similar body positions, the same person, or both. If multiple datasets carry similar data then more complex heterogeneous cross-dataset heterogeneity will arise~\cite{sanabria2021contrasgan}. Different experimental settings for cross-sensor heterogeneity vary the complexity of the data distribution heterogeneity, however, it is a very challenging task compared to cross-person and cross-position heterogeneity.~\cite{wang2018stratified} evaluates the proposed approach for a limited transfer learning setting whereas comparison between~\cite{sanabria2021contrasgan}and~\cite{zhou2020xhar} is not straightforward due to the differences in the evaluation datasets. However, ~\cite{sanabria2021contrasgan} performed better in terms of the reported F1-score and the considered activity set in~\cite{sanabria2021contrasgan}(9-10 activities) and ~\cite{zhou2020xhar}(5 activities). In Case 3, the generative unsupervised approach~\cite{akbari2019transferring} outperforms the semi-supervised approach~\cite{khan2018scaling} on the cross-device heterogeneity evaluation on the HHAR dataset. In case 4,~\cite{sanabria2020unsupervised} reported a detailed cross-sensor heterogeneity evaluation that considers 6 different datasets and achieves a 78.5\% F-1 score. In general, it is seen that generative approaches outperform other techniques in tackling cross-sensor or relevant heterogeneity. Apart from transfer learning approaches, there are approaches that deploy data augmentation, and active learning to tackle the sensor heterogeneity~\cite{mathur2018using, gudur2019activeharnet}. By increasing the size of the training dataset, data augmentation~\cite{um2017data} can help reduce the impact of sensor heterogeneity by increasing the diversity of the training data. Apart from these heterogeneous scenarios,~\cite{chen2020deep} mention the \textsc{``Concept drift''} which is a phenomenon that occurs when the underlying probability distribution of the data changes over time. In the context of wearables, concept drift can occur when the sensor readings change due to factors such as device wear and tear, changes in the environment, or changes in the wearer's behavior, which can lead to a mismatch between the training data and the test data.~\cite{abdallah2015adaptive, rokni2018autonomous} focused on activity recognition under evolving data stream, however, we note that a longitudinal dataset is a significant component to validate such methodologies that aim to can withstand the concept drift.

\section{Potential Future Directions}
\label{guidelines}

In this section, we iterate several unexplored directions that can serve as a potential guideline in sensor-based domain adaptation.

\subsection{Comprehensive Empirical Study}
\label{sub:evaluation_guidelines}
We observe that different literatures follow different experimental designs in the evaluation process. Often, the evaluation is based on selective settings of cross-person and cross-position heterogeneity instead of an exhaustive one. To elaborate, the Opportunity dataset is collected from 4 users, where each user carries 5 devices at different body positions. If we consider only cross-position heterogeneity, there would be 80 domain adaptation settings from 4 users in total (for each person, there are 20 possible combinations of positions, with different positions as the source and target domain). Similarly, for cross-person heterogeneity for a fixed body position, there are a total of 60 possible domain adaptation combinations. However, we observe limited evaluation reports in the published literature. In addition, we observe disagreement in other aspects of the evaluation process: adapting different experimental designs (splitting user groups into two - one is used as the source, the other as the target domain), different numbers of considered activities from the considered datasets~\cite{akbari2019transferring, eldele2022cotmix}, different evaluation metrics~\cite{sanabria2021contrasgan, wang2018stratified}. To bring the chaotic evaluation process into a standardized procedure, we suggest evaluating the proposed approach on three datasets among the OPPORTUNITY, PAMAP2, DSADS, MHEALTH, HHAR, Susses-Huawei, CASAS, and WISDM datasets for cross-person, cross-position, and cross-sensor heterogeneity due to their diverse number of users, positions, and sensors (devices). Regarding the experimental design, we note that if a combination of distinct user/position/sensor data is used as the source and target domain, it creates robust evaluation settings for any wearable domain adaptation approach. Finally, we recommend F1-score as the evaluation metric instead of accuracy, as it does not consider the class distribution. Due to the above-mentioned differences in the evaluation settings, it is not trivial to compare the proposed approaches.~\cite{chang2020systematic} investigated three different wearable domain adaptation methodologies (data augmentation, distance minimization, and adversarial learning) on a fixed CNN-based architecture, and~\cite{ragab2022adatime} developed a benchmarking suit for time series domain adaptation approaches. ~\cite{chang2020systematic,ragab2022adatime} contribute significantly towards the comparive study of various approaches. However,~\cite{chang2020systematic} acknowledged the use of a single framework, and~\cite{ragab2022adatime} considered approaches mainly developed for computer vision domain adaptation problems; adapting for the time series domain approach requires substantial network and learning-related hyper-parameter tuning, which is another challenge in itself. According to the literature, 20–30\% labeled target domain data improves the model's performance significantly. We envision that the benchmarking efforts on the unsupervised approaches can be further improved by incorporating our recommendation on the evaluation process and extending the baseline adaptation methods by incorporating distance and correlation-based feature alignment, adversarial and GAN-based approaches, and normalization-based approaches. Such efforts would be a significant contribution toward better understanding the strengths and weaknesses of each approach and building upon existing work in the field.

\subsection{Real-time Domain Adaptation and System Development}
Traditional domain adaptation approaches the availability of the target domain data prior to any domain adaptation process. Whereas there are certain circumstances when the target domain data arrive sequentially at large volume and it is unfeasible to store all the data. Given such scenario, it is important to perform domain adaptation in a real-time manner. To the best of our knowledge, the performance and aspects of wearable domain adaptation approaches under such circumstances have not been well investigated. The following challenging aspects of the proposed approaches need to be broadly studied: the requirement for low latency and high computational efficiency, the amount of required labeled and unlabeled target domain data, feasibility of the necessity of the device data storage, and maintaining the performance on the source and target tasks. Regarding the computation efficiency, batch normalization~\cite{ioffe2015batch} is known for faster training and quick network convergence~\cite{goodfellow2016deep}. Such parameter-free batch normalization-based approaches~\cite{li2016revisiting, carlucci2017autodial} can be investigated to develop novel real-time domain adaptation approaches. ~\cite{mazankiewicz2020incremental} adopted AdaBN~\cite{li2016revisiting} in developing an incremental domain adaptation approach.~\cite{mazankiewicz2020incremental} assumes that no samples from the target user are available in advance, but they arrive sequentially, which is substantially different than the traditional domain adaptation assumption. Thus, the authors traded off adapting to new and forgetting old information, and the proposed unsupervised approach does not involve any label information from the target user in the adaptation process.~\cite{mazankiewicz2020incremental} performs the domain adaptation task on-the-fly, which is very crucial in certain circumstances, for example when the unlabeled data can not be stored. Existing incremental learning~\cite{wulfmeier2018incremental, venkataramani2018towards} and continual learning-based~\cite{de2019continual, parisi2019continual} approaches can be explored to fullfill the requirement of a real-time systems. Further, to accommodate the deep models in the resource constrained environments, the investigation of the model compressing techniques~\cite{howard2017mobilenets, iandola2016squeezenet, lane2017squeezing, lee2019mobisr, cheng2017survey, choudhary2020comprehensive} with the domain adaptation techniques can be a significant contribution to the WHAR research and help take a step towards developing a system that is close to the benefit for the human beings through various applications.

\subsection{Investigation on Pseudo-label Generation}
In the literature review, the semi-supervised approaches acquire labeled target domain data in three approaches - 1) prior availability of the labeled target domain data, 2) asking the oracle for the label information for the most informative target domain data samples, 3) training a model with source domain data and leveraging the trained model to predict the label information of the target domain data. Indeed, label information aids in the alignment of the marginal and conditional distributions. However, we note that the pseudo-labeled-based co-learning approaches relatively underperform compared to the unsupervised approaches~\cite{wang2018stratified}. We previously mentioned that the incorrect pseudo-labeling might trigger negative transfer learning. Along with the pseudo-label generation, it is important to ensure its correctness. Novel pseudo-labeling techniques for wearables will definitely help reduce the overhead of asking the label information from the oracle or the end-user while improving the performance.

\subsection{Novel and Scalable Architecture Development}
The majority of the proposed wearable domain adaptation approaches have several algorithmic limitations in their assumptions, which consider a single labeled source domain and one target domain, and that all data domains are homogeneous. The number of considered source domains is a potential limitation of the framework regarding its scalability, as the performance of those frameworks under multiple source domains is unknown. There are several approaches to dealing with multiple source domains: 1) selecting the most relevant source domain with the target domain~\cite{chen2019cross}, 2) combining all source domain data into one~\cite{jeyakumar2019sensehar}, 3) concurrently processing all the source domains during the domain adaptation~\cite{chakma2021activity, wilson2020multi, chakma2022semi}. Concurrent processing of multiple source domains is known as multi-source domain adaptation (MSDA)~\cite{zhao2020multi, sun2015survey, guo2018multi}. MSADA assumes that the labeled training data can be collected from multiple data sources, whereas most of the current wearable domain adaptation approaches assume that source samples are collected from a single domain. As the wearable devices are used by a diverse end-user community in a diverse environment, the underlying data distributions across the collected data sources are heterogeneous. Indeed, MSDA is a very challenging task due to the simultaneous consideration of multiple heterogeneous distributions, but at the same time, it provides an opportunity to leverage the target domain-relevant domain knowledge shared across multiple source domains. The assumption of homogeneous features across all the domains is another limitation and opportunity to improve for future research work~\cite{bhalla2021imu2doppler, chatterjee2020laso, chen2019synergistic, ma2019deep, jaritz2020xmuda, tan2017distant}. We note that an inherent challenge in both of the improvements over the existing approaches is the early detection of transfer learning feasibility among multiple domains during the adaptation process. Because attempting to accomplish transfer learning between (among) non-related domains would cause the \textsc{``catastrophic forgetting''} and the initial non-relevancy might be caused by the data distribution heterogeneity. More investigation and study is needed to develop novel, scalable wearable domain adaptation approaches.

\subsection{Robust Dataset Collection}
Table~\ref{table:dataset} tabulates the widely used datasets in the wearable domain adaptation evaluation process. We note several limitations in the existing datasets: 1) The majority of the datasets consider the common macro activities (activities that require whole-body movements, such as running or playing a sport), and the micro activities (fine-grained movements and gestures, such as typing on a keyboard or playing an instrument) are omitted, 2) lack of multi-modal datasets with heterogeneous feature space (for example: datasets with concurrent activity data from paired data modalities such as IMU and Wi-Fi data, IMU and acoustic, IMU and radar, and IMU and image, 3) lack of a longitudinal dataset, which is essential to evaluate the \textsc{``concept drift''}~\cite{khamassi2018discussion, mehta2017concept}, and 4) missing interleaved activities such as cooking activities in the existing dataset. Fulfilling the existing dataset limitations would create a significant opportunity to understand the synergy between the data modalities and to explore the solution ideas discussed in the literature~\cite{khamassi2018discussion, mehta2017concept}.

\section{Conclusion}
\label{conclusion}

% Deep learning-based approaches outperform the traditional machine learning approaches using automatic feature extracting capability and the feasibility of transferring the learned knowledge from one domain to another. Even such deep learning approaches perform sub-optimal while dealing with the data distribution heterogeneity, and here, domain adaptation techniques aid alleviating such challenges. In this survey, we explain and categorize the IMU-based domain adaptation approaches that help understand the current literature progress and discover potential gaps. Based on our discussion, we further discuss the research directions and point out existing drawbacks from an IMU-based domain adaptation perspective. We hope that this survey gives the reader a birds' eye view of the current progress and helps to take on new challenges to make progress in IMU-based human activity recognition.

Deep learning-based approaches outperform the traditional machine learning approaches using automatic feature extracting capability and the feasibility of transferring the learned knowledge from one domain to another. However, in the presence of data distribution heterogeneity between training and test data samples, machine learning approaches perform poorly. In this survey, we explain a transfer learning technique called domain adaptation that focuses on alleviating data distribution heterogeneity and tabulates various types of domain adaptation problems. We concentrated on the heterogeneity observed in wearable IMU-based human activity recognition. We discuss the current literature based on the methodology existing approaches follow, the observed heterogeneity, and we concisely tabulate them. In the existing literature, we note that adversarial and generative adversarial-based approaches perform superiorly among the unsupervised approaches. Among the semi-supervised approaches, the pseudo-labeled-based approaches perform relatively lower than other semi-supervised methodologies (we speculate that it is because of ``confirmation bias'' of the pseudo-labeled data samples). However, we observe that the existing approaches lack commonality in the method evaluation process (see the ``Heterogeneity'' and ``Dataset'' columns from Table~\ref{tab:summary-all}). Due to differences in the training mechanism, number of considered heterogeneity, evaluation dataset, and considered activities, it is still difficult to compare the approaches comprehensively. We have developed a guideline based on our observations to aid in the development of a comprehensive empirical study of wearable domain adaptation approaches. In addition, we have listed several potential future guidelines - developing the real-time domain adaptation techniques for faster adaptation in a real-life scenarios, investigating more on to self-supervised learning technique such as correctly generating pseudo labels for the unlabeled data samples, extending current approaches and develop novel scalable architecture such that multiple heterogeneous datasets can be leveraged for the domain adaptation purpose, preparing a robust dataset such that it can be experimented for multiple heterogeneity evaluation. From the discussed literature, we also note that the various types of domain adaptation variations tabulated in Table~\ref{tab:da_type_definition} have not yet been well explored. We hope that this survey will provide the reader with a comprehensive overview of the current progress on wearable IMU-based domain adaptation approaches and provide guidelines for potential future developments.

{\footnotesize
\bibliography{Bibtex/Introduction-bibtex, Bibtex/DA-bibtex}}

\end{document}